\documentclass{article}

\usepackage{caption}
\usepackage{subcaption}

\usepackage{placeins}
\usepackage{graphicx}

\usepackage{algorithm}
\usepackage{algorithmic}

\usepackage{rotating}
\usepackage{ragged2e}
\usepackage{amsmath}
\usepackage{amsfonts}
\usepackage{amssymb}
\usepackage{cite}

\usepackage{longtable}
\usepackage{multicol}
\usepackage{float}

\newtheorem{thm}{Theorem}
\newtheorem{Definition}[thm]{Definition}

\date{}
\usepackage{balance}

\usepackage{caption}

\begin{document}

\title{Comparison Issues in Large Graphs: State of the Art and Future Directions}



\author{Hamida Seba*  Sofiane Lagraa** Elsen Ronando* \\
*Universit\'{e} de Lyon, CNRS, Universit\'{e} Lyon 1\\
 LIRIS, UMR5205, F-69622 Lyon, France.\\
hamida.seba@univ-lyon1.fr\\
**Universit\'{e} Grenoble Alpes\\
 CNRS, TIMA, LIG, F-38031 Grenoble, France. \\
}

\maketitle
\begin{abstract}
Graph comparison is fundamentally important for many applications such as the analysis of social networks and biological data  and has been a significant research area in the pattern recognition and pattern analysis domains.
Nowadays, the graphs are large, they may have billions of nodes and edges. Comparison issues in such huge graphs are a challenging research problem.

In this paper, we survey the research advances of comparison problems in large graphs. 
We review graph comparison and pattern matching approaches that focus on large graphs.
We categorize the existing approaches into three classes: partition-based approaches, search space based approaches and summary based approaches. All the existing algorithms  in these approaches are described in detail and analyzed according to multiple metrics such as time complexity, type of graphs or comparison concept. Finally, we identify directions for future research.

\end{abstract}

\section{Introduction}
\label{sec:introduction}
Comparing objects is one of the most frequently encountered tasks in computing: information retrieval, pattern recognition, biology, computer vision, etc. A comparison problem occurs whenever an object or a piece of it needs to be mapped to another object or part of it.
Graphs are an attractive representation and modeling tool since they allow simple, intuitive and flexible representations of complex and interacting objects. Consequently, object comparison leads generally to a problem of graph comparison.
Although significant progress has been made in graph comparison and related areas such as graph/subgraph isomorphism, pattern matching, etc., the recent explosion of the size of data generated and manipulated daily by applications and human activities has given rise to the big graph data challenge. In fact, real-world graphs are large and even huge, i.e., thousands, millions and even billions nodes and edges. Social networks, web graphs and protein interaction graphs are some examples. For these graphs, existing solutions for graph analysis, mining, visualization, etc., do not scale at all. These algorithms must be revisited or even re-invented.

Traditional graph comparison approaches are generally classified into two categories: exact approaches and inexact approaches. Exact approaches, such as graph isomorphism, sub-graph isomorphism and the maximum common subgraph, aim to find out if an exact mapping between the vertices and the edges of the compared graphs or subgraphs is possible
\cite{Ullmann76,Mckay1981,Bunke2011,Conte2004}.

Inexact graph comparison aims generally to compute a distance between the compared graphs.  This distance measures how much these graphs are similar and helps to deal with the errors and the noise that is inevitably introduced during the process needed to model objects by graphs. Inexact graph comparison is also useful for search/rank based applications where a distance between the compared objects is needed. In some applications, graph similarity measures are intended to compute relatively suboptimal distances \cite{Conte2004} that are compensated by a large reduction of the computational complexity of the comparison process.
Several graph similarity measures have been proposed in the literature and several approaches have been used including genetic algorithms \cite{Khoo2001,Suganthan2002}, neural networks \cite{Micheli2009}, the theory of probability \cite{Myers2000,William1995}, clustering techniques \cite{Hancock2001,Alberto2000}, spectral methods \cite{Umeyama1988,hancock2007}, decision trees \cite{Messmer1999,Messmer1995}, etc.
We refer the reader to \cite{Brian1939, Bunke1997,Conte2004, Bunke2011, Vento2013} for more exhaustive surveys.
In order to cope with large graphs, new techniques, concepts and approaches have been proposed recently for performing  graph comparison. Thus, in this paper we focus mainly on the solutions designed for large graphs.

The aim of this paper is to provide a survey of recent and current development of graph comparison and pattern matching approaches on large graphs. We describe and analyze in detail the existing approaches and we categorize them into different classes. We also highlight the advantages, disadvantages and the differences between the approaches and identify direction for future research.

The rest of the paper is organized as follows:  Section~\ref{ProblemDef} presents the problem definition and preliminaries. Section~\ref{Approaches} presents the different approaches that we have categorized,  analyzed and described in detail in order to compare them and to show their advantages and disadvantages.  A summary of these approaches is presented and some important problems of graph comparison and pattern matching deserving further research are proposed in Section~\ref{sec:Discussion}. Section~\ref{sec:conclusion} concludes the paper.

\section{Problem Definition and Basics}
\label{ProblemDef}
In this section we present some basic definitions related to graphs and their comparison problems. We rely mainly on the terminology used in \cite{Basu2006,Fan2010}. So, all the definitions below are adapted from \cite{Basu2006,Fan2010}.
\begin{Definition} A graph $G$ is a 4-tuple $G =(V, E,f_V,f_E)$, where $V$ is a set of nodes (also
called vertices), $E\subseteq V\times V$ is a set of edges connecting the nodes, $f_V: V\rightarrow\Sigma_V$ and $f_E: E\rightarrow\Sigma_E$ are functions labeling the nodes and the edges respectively where $\Sigma_V$ and $\Sigma_E$ are the the sets of labels that can appear on the nodes and edges, respectively.
\end{Definition}

When omitting $f_E$ in the definition of $G$, we mean that $\Sigma_E$ is an empty set and the graph is not edge labeled. So, when there is no ambiguity, the notation $G =(V,E)$ defines vertex labeled graph. We will also use the terms vertex and node interchangeably in all this document. \\

The edges of a graph may have a direction associated with them. In this case, the graph is directed.\\
Generally, the number of vertices of a graph is called the order of the graph and the number of its edges is called the size of the graph.

A graph that is contained in another graph is called a subgraph and is defined as follows:
\begin{Definition}
A graph $G_1 = (V_1,E_1, f_{V_1}, f_{E_1})$ is a subgraph of a graph $G_2 = (V_2,E_2, f_{V_2}, f_{E_2})$, denoted $G_1 \subseteq G2$, if $V_1\subseteq V_2$, $E_1\subseteq E_2\cap(V_1 \times V_1)$,
$f_{V_1}(x)=f_{V_2}(x) \forall x \in V_1$,
and $f_{E_1}((x, y))=f_{E_2}((x, y))$ $\forall(x,y)\in E_1$.
\end{Definition}

The distance between two nodes $u$ and $v$ in a graph $G$, denoted by $dist(u,v)$, is the length
of the shortest undirected path from $u$ to $v$ in $G$. The diameter of a connected graph $G$, denoted by $d_G$, is the longest shortest distance of all pairs of nodes in $G$, i.e., $d_G = max(dist(u, v))$ for all nodes $u$, $v$ in $G$.
The eccentricity of a vertex in a graph is its maximum distance from any other vertex in the graph. The vertices of the graph with the minimum eccentricity are the centers of the graph, and the value of their eccentricity is the radius of the graph. The maximum value of eccentricity equals to the diameter of the graph .

Several applications that use graphs as a modeling tool such as pattern recognition, information retrieval, mining, etc., need to compare graphs. Graph comparison, also called graph matching, has been subject of several studies and surveys such as \cite{Conte2004}, \cite{Gallagher2006}, \cite{Aggarwal2010} and \cite{Vento2013}. Graph comparison approaches are generally classified into two categories: exact approaches and inexact or fault-tolerant approaches. Exact approaches refer to the methods used to find out if two graphs are the same \cite{Ullmann76,Mckay1981,Bunke2011,Conte2004}. This means that we look for graph isomorphism.

Fault-tolerant graph comparison aims generally to compute a distance between the compared graphs.  This distance measures how much these graphs are similar and is motivated mainly by three situations:
\begin{itemize}
\item the process of modeling objects by graphs may be subject to noise and distortions. This means that a modeling process executed twice on the same object may return two slightly different graphs.
The different stages of image encoding is perhaps the most illustrative example of such noise that graph comparison must deal with \cite{Bunke2011}.
\item search/rank based applications such in database query processing or web search based applications need to compute a distance between the compared objects in order to rank the top-$k$ results \cite{Tong2007,Cheng2013}.
\item In some applications, graph similarity measures are intended to compute relatively suboptimal distances \cite{Conte2004} that are compensated by a large reduction of the computational complexity of the comparison process.
\end{itemize}
In both approaches and depending on the application, we need either to compare two whole graphs or a query graph with a large graph. According to this, graph comparison methods can be classified into two categories: graph similarity measures and graph pattern matching methods.

\subsection{Graph similarity/dissimilarity measures}
The aim of similarity/dissimilarity measures is to quantify the degree of resemblance between two graphs. The strongest similarity degree is graph "equality", called graph isomorphism and defined as follows:
\begin{Definition} A graph $G_1 = (V_1,E_1, f_{V_1}, f_{E_1})$ and a graph $G_2 = (V_2,E_2, f_{V_2}, f_{E_2})$ are said to be isomorphic, denoted $G_1 \cong G2$, if there exists a bijective function $h : V_1\rightarrow V_2$ such that the following conditions are met:
\begin{enumerate}
\item $\forall x \in V_1: f_{V_1}(x) = f_{V_2}(h(x))$
\item $\forall(x,y)\in E_1: (h(x), h(y))\in E_2$ and $f_{E_1}((x,y))=f_{E_2}((h(x), h(y)))$
\item $\forall(h(x),h(y))\in E_2: (x,y)\in E_1$ and $f_{E_2}((h(x),h(y)))=f_{E_1}((x, y))$
\end{enumerate}
\end{Definition}

Several relaxed approaches, i.e., "fault-tolerant graph comparison", are also proposed. They are useful for search/rank based applications where a distance between the compared objects is needed. In some applications, graph similarity measures are intended to compute relatively suboptimal distances \cite{Conte2004} that are compensated by a large reduction of the computational complexity of the comparison process.

Several graph similarity measures have been proposed in the literature and several approaches have been used including genetic algorithms \cite{Khoo2001,Suganthan2002}, neural networks \cite{Micheli2009}, the theory of probability \cite{Myers2000,William1995}, clustering techniques \cite{Hancock2001,Alberto2000}, spectral methods \cite{Umeyama1988,hancock2007}, decision trees \cite{Messmer1999,Messmer1995}, etc.
We refer the reader to \cite{Brian1939,Bunke1997,Conte2004,Bunke2011,Vento2013} for more exhaustive surveys.
Some of the existing approaches try to extend to graphs some of the properties defined in metric spaces.
\begin{Definition}
A metric space is an ordered pair $(M,d)$ where $M$ is a set and $d$ is a metric on $M$, i.e., a function 
\begin{itemize}
\item $d(x,y)\geq 0$ (non-negativity),
\item $d(x,y)=0$ iff $x=y$ (uniqueness),
\item $d(x,y)=d(y,x)$ (symmetry) and
\item $d(x,z)\leq d(x,y)+d(y,z)$ (triangle inequality).
     \end{itemize}
\end{Definition}

\begin{figure*}[tbph]
\centering
\includegraphics [scale=0.4]{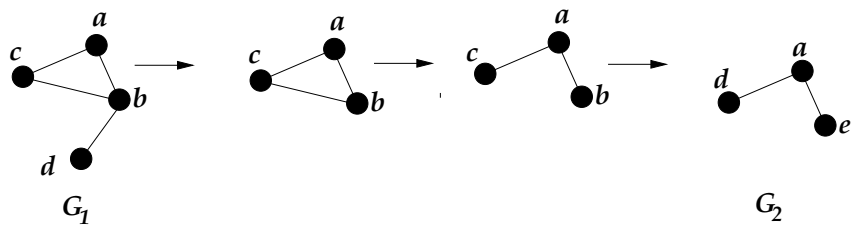}
\caption{\label{fig:Example-of-edit}Example of edit operations \cite{Lagraa2014}.}
\end{figure*}

Perhaps, the most referenced metric is edit distance which defines the similarity of graphs by the minimum costing sequence of edit operations that convert one graph into the other \cite{Sanfeliu1983,Bunke1983}. An edit operation is either an insertion, a suppression or a re-labeling of a vertex or an edge in the graph. A cost function associates a cost to each edit operation.
Figure \ref{fig:Example-of-edit} shows an example of edit
operations that are necessary to get the graph $G_{2}$ from $G_{1}$ with the suppression of two edges and a vertex and the relabeling of two vertices.

Graph edit distance is a flexible 
graph similarity measure which is applicable to various kinds of graphs \cite{Bunke1983,Neuhaus2004,Sanfeliu1983,Ambauen2003,Robles2005}. It also defines a common theoretical framework that allows comparing different approaches of graph comparison. In fact, Bunke showed in \cite{Bunke1997b} that under a particular cost function, graph edit distance computation is equivalent to the maximum common subgraph problem. In \cite{Bunke1999}, the same author shows that the graph isomorphism and subgraph isomorphism problems can be reduced to graph edit distance. However, computing graph edit distance suffers from two main drawbacks:
\begin{enumerate}
\item A high computational complexity. The problem of computing graph edit distance is NP-hard in general \cite{Zeng2009}. The most known method for computing the exact value of graph edit distance is based on $A^*$ \cite{Hart1968} which is a best first search algorithm where the search space is organized as a tree. The root of the tree is the starting point of the algorithm. The internal vertices correspond to partial solutions and leaves represent complete solutions.
\item The difficulty related to defining cost functions \cite{Neuhaus2007}.
\end{enumerate}
The first drawback motivated several approximating solutions to compute graph edit distance. A comprehensive survey on graph edit distance and the approaches proposed to compute it can be found in \cite{Gao2010}.
To overcome the second drawback and avoid the definition of edit costs, similarity measures that do not use edit operations are also proposed. In \cite{Bunke1998}, the authors propose a graph distance measure that is based on the maximal common subgraph of two graphs and prove that it is a metric, i.e., the measure satisfies the four properties of a usual metric namely: non-negativity, uniqueness, symmetry and triangle inequality. However, computing the maximal common subgraph of two graphs has a high computational complexity \cite{Bunke1998}. For this reason, Raymond et al. \cite{Raymond2002} propose a modified version of the measure defined in \cite{Bunke1998} where an initial screening process determines whether it is possible for the measure of similarity between the two graphs to exceed a minimum
threshold for which it is acceptable to compute the maximum common subgraph. This screening process is based on computing graph invariants.  Graph invariants have been efficiently used to solve the graph comparison problem in general and the graph isomorphism problem in particular. They are used for example in Nauty \cite{Mckay1981} which is one of the most efficient algorithm for graph and subgraph isomorphism testing. A vertex invariant, for example, is a number $i(v)$ assigned to a vertex $v$ such that if there is
an isomorphism that maps $v$ to $v'$ then $i(v)=i(v')$. Examples
of invariants are the degree of a vertex, the number of cliques of
size $k$ that contain the vertex, the number of vertices at a given
distance from the vertex, etc.
Graph invariants are also the basis of graph probing \cite{Lopresti2001} where a distance between two graphs is defined as the norm of their probes. Each graph probe is a vector of graph invariants.

In \cite{Wallis2001}, the distance metric based on the maximum common subgraph defined in \cite{Bunke1998} is extended by a proposal to define the problem size with the union of the two compared graphs rather than the larger of the two graphs used in \cite{Bunke1998}.

In \cite{Xiao08}, the authors show that we can evaluate graph distance with a high degree of precision by considering complex graph sub-structures in the distance. In fact, in some applications such as analysis of protein interaction graphs, some sub-structures of these graphs represent certain functional modules of cells or
organisms. Hence, comparing these graphs in terms of substructure
information is biologically meaningful \cite{Xiao08}. The authors defined a new metric based on the concept of \textit{Structure Abundance Vector}. Each element of a Structure Abundance Vector of a graph $G$ contains the size of an occurrence of a predefined sub-structure in $G$. The \textit{Structure Abundance Vector} is a generalization of the concept of graph invariants. \\
 More recently, kernel based similarity measures are also proposed \cite{Haussler1999,Gartner2003,Borgwardt2005,Neuhaus2006a,Neuhaus2006c,Bunke2011}. The main idea is also to define similarity of graphs based on the similarity of substructures of these graphs.\\

\subsection{Subgraph/Pattern matching}
Given two graphs $Q$ and $G$, the graph pattern matching problem is to find all subgraphs of $G$ that match $Q$. In other words, find all the embeddings of $Q$ in $G$. Generally, $Q$ is called the query graph or simply pattern and $G$ is large compared to $Q$. The exact version of graph pattern matching is called Subgraph isomorphism and is defined as follows:
\begin{Definition} A graph $Q = (V_Q,E_Q, f_{V_Q}, f_{E_Q})$ is subgraph isomorphic to a graph $G = (V_G,E_G, f_{V_G}, f_{E_G})$ if there exists a subgraph $G'$ of $G$ such that $Q$ and $G'$ are isomorphic.
\end{Definition}

Subgraph isomorphism is an NP-complete problem \cite{Garey1979}. The most known methods to enumerate the subgraphs of $G$ that are isomorphic to a query $Q$ are based on exploring search spaces. With these approaches, the number of possible matchings to be checked increases combinatorially with the number of nodes in the graphs. Even with the help of pruning methods that reduces the size of the search space \cite{Ullmann76,Cordella2004}, these methods for subgraph isomorphism checking remain impractical for large graphs such as social networks. Furthermore, these graphs are directed, i.e., $(u,v)$ and $(v,u)$ denote different edges, and edge labeled. Moreover, the considered graph patterns are not simple graphs. A pattern in this kind of applications is a "regular expression"-like graph where a node is labeled by a search conditions which specifies a set of possible values for the node and the edge.  In \cite{Cheng2008}, an edge in query graph, is a directed edge and does not correspond to a direct edge between two nodes but to some reachability condition that means that the endpoint of the edge is reachable from the source node of the edge. This idea was extended in \cite{Zou2009} by the introduction of a bound $\delta$ such that if there is an edge between two nodes in the query, these nodes are mapped into the data graph to two nodes reachable within $\delta$ edges, i.e., the shortest path between the two nodes is at most $\delta$. More recently, \cite{} introduces "regular expression"-like graph patterns that combine the concept of bounded edges of~\cite{Zou2009} with the power of regular expressions for defining the possible value taken by the labels of the nodes.

Consequently, relaxed approaches that achieve a better time complexity and that are more adapted to these pattern-based applications are proposed. In this context, \textit{Graph simulation} \cite{Milner1989,Henzinger1995} receives an increasing interest specially for social network analysis. Graph simulation is defined as follows:
\begin{Definition}
 A  pattern $Q = (V_{Q},E_{Q},f_{V_Q},f_{E_Q})$ matches a directed graph $G = (V_G,E_G,f_{V_G},f_{E_G})$ via simulation, denoted by $Q\trianglelefteq G$, if there exists a binary relation $S\subseteq V_{Q}\times V_G$ such that:
\begin{enumerate}
\item for each $u\in V_{Q}$, there exists $v\in V_G$ such that $(u,v)\in S$;
\item for each $(u,v)\in S$, we have
\begin{enumerate}
\item $f_{V_Q}(u)=f_{V_G}(v)$;
\item for each edge $(u,u')\in E_Q$ there is an edge $(v,v')\in E_G$ such that $(u',v')\in S.$
\end{enumerate}
\end{enumerate}
\end{Definition}

\begin{figure}[tbph]
\centering
\includegraphics [scale=0.4]{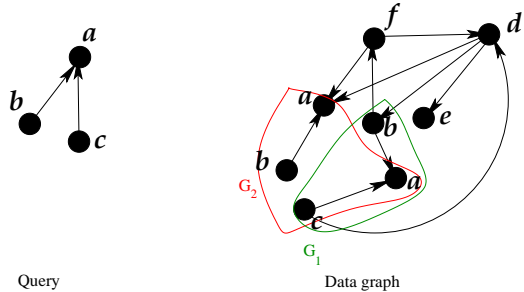}
\caption{Subgraph isomorphism Vs graph simulation.}
\label{Fig-iso-sim}
\end{figure}

The graph that corresponds to simulation $S$ is called the \textit{match graph} and is defined as follows:
\begin{Definition}
Let $Q = (V_{Q},E_{Q},f_{V_Q},f_{E_Q})$ be a query graph that matches a data graph $G = (V_G,E_G,f_{V_G},f_{E_G})$ via simulation  $S\subseteq V_{Q}\times V_G$.
 The match graph that corresponds to $S$ is a subgraph $G_S$ of $G$ such that $G_S=(V_S,E_S)$, in which (1) a node $v\in V_S$ iff it is in $S$, and (2) an edge $(v,v')\in E_S$ iff there exists an edge $(u,u')\in E_Q$  with $(u,v)\in S$ and $(u',v')\in S$.
\end{Definition}
Contrarily to isomorphism, when two graphs $G_1$ and $G_2$  match by simulation, a node of one graph may be mapped to several  nodes in the second graph.
In Figure \ref{Fig-iso-sim}, the query graph is isomorphic to subgraph $G_1$ but it matches by simulation subgraphs $G_1$ and $G_2$. We note also that $G_2$ is not connected. \\

Note that a quadratic time algorithm for graph simulation is proposed in \cite{Henzinger1995}.

\section{Approaches}
\label{Approaches}

\begin{figure*}[tbph]

%
%
\centering
\setlength{\unitlength}{1776sp}%
\begingroup\makeatletter\ifx\SetFigFont\undefined%
\gdef\SetFigFont#1#2#3#4#5{%
  \reset@font\fontsize{#1}{#2pt}%
  \fontfamily{#3}\fontseries{#4}\fontshape{#5}%
  \selectfont}%
\fi\endgroup%

\begin{picture}(15134,8363)(350,-7611)
\thicklines{\put(7226,-1786){\framebox(5150,930){}}
}%
{\put(4703,-2911){\framebox(1028,660){}}
}%
{\put(3297,-2881){\framebox(1028,660){}}
}%
{\put(1703,-2911){\framebox(1028,660){}}
}%
{\put(383,-2836){\framebox(1028,645){}}
}%
{\put(10366,-3151){\framebox(2090,885){}}
}%
{\put(7351,-3136){\framebox(2400,900){}}
}%
{\put(12556,-1771){\framebox(2995,893){}}
}%
{\put(5581,-241){\framebox(5280,960){}}
}%
{\put(1906,-1726){\framebox(3020,900){}}
}%
{\put(5948,-2926){\framebox(1028,660){}}
}%
{\put(7771,-3136){\line( 0,-1){1230}}
}%
{\put(11000,-3129){\line( 0,-1){2677}}
}%
{\put(6260,-2919){\line( 0,-1){592}}
}%
{\put(4891,-2911){\line( 0,-1){2415}}
}%
{\put(3425,-3496){\line( 1, 0){270}}
}%
{\put(3440,-2904){\line( 0,-1){592}}
}%
{\put(1880,-2889){\line( 0,-1){4597}}
}%
{\put(436,-3443){\line( 1, 0){195}}
}%
{\put(436,-2851){\line( 0,-1){570}}
}%
{\put(12950,-1734){\line( 0,-1){2062}}
}%
{\put(7936,-211){\line( 0,-1){375}}
\put(7936,-586){\line(-1, 0){4410}}
\put(3526,-571){\line( 0,-1){240}}
}%
{\put(7936,-571){\line( 1, 0){6090}}
\put(14026,-586){\line( 0,-1){270}}
}%
{\put(9916,-586){\line( 0,-1){240}}
}%
{\put(3286,-1741){\line( 0,-1){240}}
\put(3286,-1981){\line(-1, 0){2340}}
\put(946,-1981){\line( 0,-1){165}}
}%
{\put(3286,-1981){\line( 1, 0){3150}}
\put(6436,-2011){\line( 0,-1){255}}
}%
{\put(2206,-1981){\line( 0,-1){255}}
\multiput(2206,-2236)(15.00000,15.00000){2}{\makebox(14.8148,22.2222){\small.}}
}%
{\put(3796,-1981){\line( 0,-1){225}}
}%
{\put(5176,-1996){\line( 0,-1){240}}
}%
{\put(9931,-1786){\line( 0,-1){240}}
\put(9931,-2026){\line(-1, 0){1530}}
\put(8401,-2026){\line( 0,-1){210}}
}%
{\put(9916,-2011){\line( 1, 0){1455}}
\put(11371,-2011){\line( 0,-1){240}}
}%
{\put(1869,-3440){\line( 1, 0){195}}
}%
{\put(1869,-4100){\line( 1, 0){195}}
}%
{\put(1869,-4790){\line( 1, 0){195}}
}%
{\put(1869,-5480){\line( 1, 0){195}}
}%
{\put(1854,-6155){\line( 1, 0){195}}
}%
{\put(1854,-6830){\line( 1, 0){195}}
}%
{\put(1854,-7475){\line( 1, 0){195}}
}%
{\put(4884,-3800){\line( 1, 0){195}}
}%
{\put(4869,-4520){\line( 1, 0){195}}
}%
{\put(4869,-5330){\line( 1, 0){195}}
}%
{\put(7749,-4370){\line( 1, 0){195}}
}%
{\put(6249,-3515){\line( 1, 0){195}}
}%
{\put(10974,-3710){\line( 1, 0){195}}
}%
{\put(11004,-4370){\line( 1, 0){195}}
}%
{\put(10989,-5045){\line( 1, 0){195}}
}%
{\put(10989,-5810){\line( 1, 0){195}}
}%
{\put(12954,-2330){\line( 1, 0){195}}
}%
{\put(12939,-3080){\line( 1, 0){195}}
}%
{\put(12954,-3815){\line( 1, 0){195}}
}%
\put(661,-3489){\makebox(0,0)[lb]{\smash{{\SetFigFont{8}{9.6}{\rmdefault}{\bfdefault}{\updefault}{\cite{Shapiro1985}}%
}}}}
\put(2120,-3527){\makebox(0,0)[lb]{\smash{{\SetFigFont{8}{9.6}{\rmdefault}{\bfdefault}{\updefault}{\cite{Eshera1984a}}%
}}}}
\put(3740,-3542){\makebox(0,0)[lb]{\smash{{\SetFigFont{8}{9.6}{\rmdefault}{\bfdefault}{\updefault}{\cite{Zhao2012}}%
}}}}
\put(5225,-3872){\makebox(0,0)[lb]{\smash{{\SetFigFont{8}{9.6}{\rmdefault}{\bfdefault}{\updefault}{\cite{Wang2012}}%
}}}}
\put(6511,-3541){\makebox(0,0)[lb]{\smash{{\SetFigFont{8}{9.6}{\rmdefault}{\bfdefault}{\updefault}{\cite{Zhao2013}}%
}}}}
\put(11330,-3767){\makebox(0,0)[lb]{\smash{{\SetFigFont{8}{9.6}{\rmdefault}{\bfdefault}{\updefault}{\cite{Fan2010}}%
}}}}
\put(8073,-4427){\makebox(0,0)[lb]{\smash{{\SetFigFont{8}{9.6}{\rmdefault}{\bfdefault}{\updefault}{\cite{Han2013}}%
}}}}
\put(11319,-4434){\makebox(0,0)[lb]{\smash{{\SetFigFont{8}{9.6}{\rmdefault}{\bfdefault}{\updefault}{\cite{Ma2011}}%
}}}}
\put(11304,-5139){\makebox(0,0)[lb]{\smash{{\SetFigFont{8}{9.6}{\rmdefault}{\bfdefault}{\updefault}{\cite{Fard2013}}%
}}}}
\put(11319,-5829){\makebox(0,0)[lb]{\smash{{\SetFigFont{8}{9.6}{\rmdefault}{\bfdefault}{\updefault}{\cite{Fard2014}}%
}}}}
\put(2131,-4171){\makebox(0,0)[lb]{\smash{{\SetFigFont{8}{9.6}{\rmdefault}{\bfdefault}{\updefault}{\cite{Riesen2007}}%
}}}}
\put(2094,-4854){\makebox(0,0)[lb]{\smash{{\SetFigFont{8}{9.6}{\rmdefault}{\bfdefault}{\updefault}{\cite{Zeng2009}}%
}}}}
\put(2109,-5544){\makebox(0,0)[lb]{\smash{{\SetFigFont{8}{9.6}{\rmdefault}{\bfdefault}{\updefault}{\cite{Raveaux2010}}%
}}}}
\put(2124,-6189){\makebox(0,0)[lb]{\smash{{\SetFigFont{8}{9.6}{\rmdefault}{\bfdefault}{\updefault}{\cite{Jouili2009}}%
}}}}
\put(2094,-6864){\makebox(0,0)[lb]{\smash{{\SetFigFont{8}{9.6}{\rmdefault}{\bfdefault}{\updefault}{\cite{Sun2012}}%
}}}}
\put(5214,-4539){\makebox(0,0)[lb]{\smash{{\SetFigFont{8}{9.6}{\rmdefault}{\bfdefault}{\updefault}{\cite{Khan2011}}%
}}}}
\put(5229,-5364){\makebox(0,0)[lb]{\smash{{\SetFigFont{8}{9.6}{\rmdefault}{\bfdefault}{\updefault}{\cite{Khan2013}}%
}}}}
\put(8521,-1216){\makebox(0,0)[lb]{\smash{{\SetFigFont{10}{12.0}{\rmdefault}{\bfdefault}{\updefault}{Search-space}%
}}}}
\put(1981,-1231){\makebox(0,0)[lb]{\smash{{\SetFigFont{10}{12.0}{\rmdefault}{\bfdefault}{\updefault}{Partition-based}%
}}}}
\put(6301,-61){\makebox(0,0)[lb]{\smash{{\SetFigFont{10}{12.0}{\rmdefault}{\bfdefault}{\updefault}{comparison methods}%
}}}}
\put(7291,-1636){\makebox(0,0)[lb]{\smash{{\SetFigFont{10}{12.0}{\rmdefault}{\bfdefault}{\updefault}{exploration-based methods}%
}}}}
\put(6691,359){\makebox(0,0)[lb]{\smash{{\SetFigFont{10}{12.0}{\rmdefault}{\bfdefault}{\updefault}{Large graph}%
}}}}
\put(12601,-1231){\makebox(0,0)[lb]{\smash{{\SetFigFont{10}{12.0}{\rmdefault}{\bfdefault}{\updefault}{Summary-based}%
}}}}
\put(7696,-2611){\makebox(0,0)[lb]{\smash{{\SetFigFont{9}{10.8}{\rmdefault}{\bfdefault}{\updefault}{Sub-graph}%
}}}}
\put(7501,-3046){\makebox(0,0)[lb]{\smash{{\SetFigFont{9}{10.8}{\rmdefault}{\bfdefault}{\updefault}{isomorphism}%
}}}}
\put(10861,-2626){\makebox(0,0)[lb]{\smash{{\SetFigFont{9}{10.8}{\rmdefault}{\bfdefault}{\updefault}{Graph}%
}}}}
\put(10508,-3061){\makebox(0,0)[lb]{\smash{{\SetFigFont{9}{10.8}{\rmdefault}{\bfdefault}{\updefault}{Simulation}%
}}}}
\put(518,-2584){\makebox(0,0)[lb]{\smash{{\SetFigFont{9}{10.8}{\rmdefault}{\bfdefault}{\updefault}{edge}%
}}}}
\put(1906,-2611){\makebox(0,0)[lb]{\smash{{\SetFigFont{9}{10.8}{\rmdefault}{\bfdefault}{\updefault}{star}%
}}}}
\put(3402,-2614){\makebox(0,0)[lb]{\smash{{\SetFigFont{9}{10.8}{\rmdefault}{\bfdefault}{\updefault}{path}%
}}}}
\put(4782,-2614){\makebox(0,0)[lb]{\smash{{\SetFigFont{9}{10.8}{\rmdefault}{\bfdefault}{\updefault}{tree}%
}}}}
\put(6027,-2629){\makebox(0,0)[lb]{\smash{{\SetFigFont{9}{10.8}{\rmdefault}{\bfdefault}{\updefault}{other}%
}}}}
\put(2431,-1576){\makebox(0,0)[lb]{\smash{{\SetFigFont{10}{12.0}{\rmdefault}{\bfdefault}{\updefault}{methods}%
}}}}
\put(13336,-1659){\makebox(0,0)[lb]{\smash{{\SetFigFont{10}{12.0}{\rmdefault}{\bfdefault}{\updefault}{methods}%
}}}}
\put(13235,-2417){\makebox(0,0)[lb]{\smash{{\SetFigFont{8}{9.6}{\rmdefault}{\bfdefault}{\updefault}{\cite{Fan2012b}}%
}}}}
\put(13224,-3129){\makebox(0,0)[lb]{\smash{{\SetFigFont{8}{9.6}{\rmdefault}{\bfdefault}{\updefault}{\cite{Chen2009}}%
}}}}
\put(2094,-7509){\makebox(0,0)[lb]{\smash{{\SetFigFont{8}{9.6}{\rmdefault}{\bfdefault}{\updefault}{\cite{Zheng2013}}%
}}}}
\put(13209,-3849){\makebox(0,0)[lb]{\smash{{\SetFigFont{8}{9.6}{\rmdefault}{\bfdefault}{\updefault}{\cite{Lagraa2014}}%
}}}}
\end{picture}%

\caption{\label{Fig-class-match}Classification of graph comparison approaches}
\end{figure*}
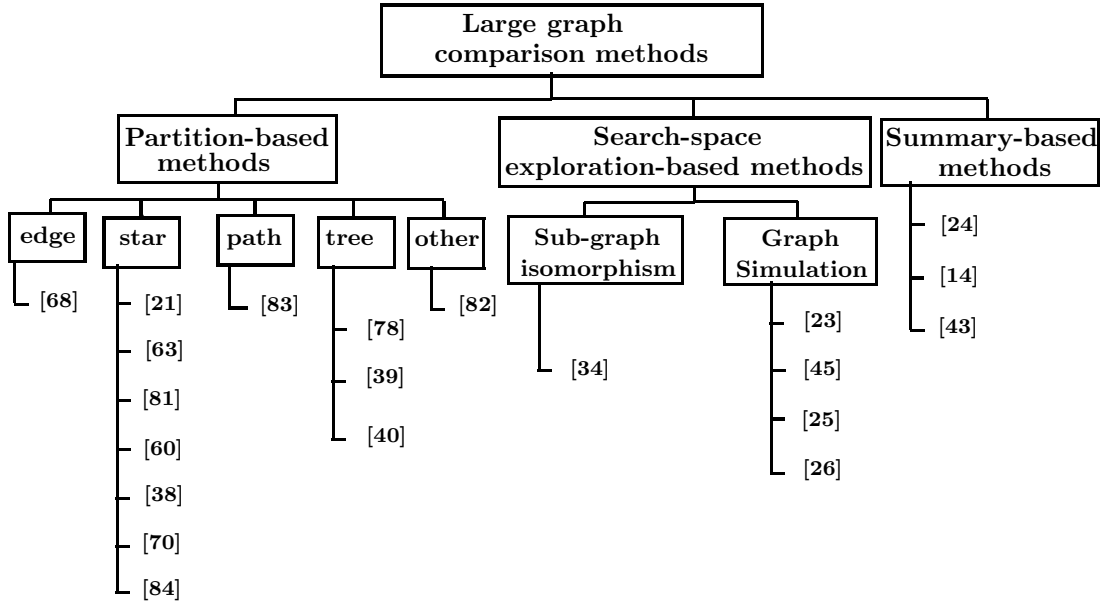

In this section, we review graph comparison and pattern matching methods that focus on large graphs. Existing approaches can be categorized into three classes: partition based approaches, search space based approaches and summary based approaches. Figure \ref{Fig-class-match} summarizes the approaches that will be reviewed in the rest of this section.

\subsection{Partition-based Approaches}
The basic idea of these approaches is to decompose graphs into sets of subgraphs and to compute the similarity between the initial graphs in function of a comparison between the obtained subgraphs. Partition-based approaches have two advantages:
\begin{enumerate}
\item They have a polynomial time complexity and thus may be suitable for large graph comparison.
\item They may highlight the existence of particular or meaningful structures within the compared graphs. These structures may enhance the accuracy of the comparison.
\end{enumerate}

The first partition-based approach dates back to the $80$s with the work of Eshera and Fu \cite{Eshera1984a,Eshera1984b}. The authors compute the edit distance between two attributed and directed graphs $G_1$ and $G_2$ in polynomial time ($O(n^2\times m^2)(n+m))$ in the worst case, where $n$ is the order of the graph and $m$ is its size). In this approach, the edit distance between $G_1$ and $G_2$ is mapped to the edit distance between their Basic Sub-Graphs called Basic Attributed Relational Graphs (BARGs) defined as follows:
\begin{Definition} \cite{Eshera1984a,Eshera1984b} A Basic attributed relational graph (BARG or Basic graph) is a graph on the form of one level tree, i.e., it consists of a root node, the branches emanating from it, and the nodes on which these branches terminate.
\end{Definition}
In other words, a BARG is a star structure composed of a root vertex, its outcoming edges and the leaves associated to these edges.
 The mapping between two sets of BARGs is achieved via the exploration of a state space organized as a directed acyclic labeled lattice. Each state of the lattice is labeled with the set of matched BARGs and denotes the reconstruction
of a subgraph from the query graph and a subgraph from the target graph as well as the matching of their respective BARGs. An edge between two states is labeled by the cost of the transition between two states. The final distance between the two graphs corresponds to the shortest costed path in the lattice. It is determined by dynamic programming.

\begin{figure}[tbph]
\centering
\includegraphics [scale=0.35]{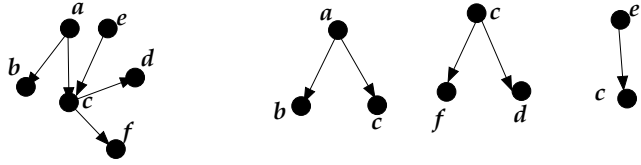}
\caption{\label{fig:Decomposition-Eschera}A graph and its decomposition into BARGs.}
\end{figure}

In \cite{Shapiro1985}, the authors consider pair of vertices and their connecting edges, called Relational Descriptions (RD)). They define a distance between two graphs based on the number of isomorphic RDs and prove that it is a metric. Given two graphs $G_1$ and $G_2$, the distance is defined by the number of RDs of $G_1$ that are not mapped to subgraphs of $G_2$ and the number of RDs of $G_2$ that are not mapped to subgraphs of $G_1$. 


In \cite{Riesen2007,Riesen2009} the authors propose a modification of the approach of Eshera and Fu \cite{Eshera1984a,Eshera1984b} that considers undirected graphs and avoids the state exploration part of the distance computation. 
In this solution, 
an optimal match between the sets of star structures, called local structures, is obtained using the Hungarian algorithm~\cite{Kuhn1955,Munkres1957}. Given a source graph $G_1$ and a target graph $G_2$, the nodes of $G_1$ are mapped to the nodes of $G_2$ using the Hungarian algorithm by defining a cost matrix that records for each vertex from $G_1$ the edit operations that are needed to transform it to each vertex of $G_2$. 

\begin{figure}[tbph]
\centering
\includegraphics [scale=0.35]{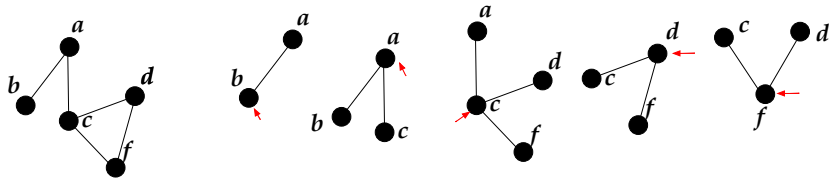}
\caption{\label{fig:Decomposition-Zeng}A graph and its star decomposition according to \cite{Zeng2009}. The arrows indicate the root of the stars.}
\end{figure}

A similar approach in presented is \cite{Zeng2009}. In this case, the graphs are also undirected. They are decomposed into multisets of stars as in \cite{Eshera1984a,Eshera1984b}. In this approach, a star structure is defined around each vertex as in \cite{Riesen2007,Riesen2009} as follows:
\begin{Definition}\cite{Zeng2009}
A star structure $s$ is an attributed,
a single-level, rooted tree which can be represented by a 3-tuple $s=(r,\mathfrak{L},\ell)$, where $r$ is the root vertex, $\mathfrak{L}$ is the set of leaves and $\ell$ is a labeling function. Edges exist between $r$ and any vertex in $\mathfrak{L}$ and no edge exists among vertices in $\mathfrak{L}$.
\end{Definition}
Figure~\ref{fig:Decomposition-Zeng} shows an example of a graph and its star decomposition.

The edit operation between two stars is defined as follows:
\begin{Definition}\cite{Zeng2009}\label{Def-Star-Zeng}
Given two star structures $s_1$ and $s_2$, the edit distance between $s_1$ and $s_2$ is:
$$ \lambda(s_1, s_2) = T(r_1, r_2) + d(\mathfrak{L}_1, \mathfrak{L}_2)$$
where
$$ T(r_1,r_2) = \left\{ \begin{array}{ll}
0 & \text{ if } \ell(r_1) = \ell(r_2), \\
1 & \text{ otherwise}.
\end{array} \right.  \\
$$
$$d(\mathfrak{L}_1, \mathfrak{L}_2) = ||\mathfrak{L}_1| - |\mathfrak{L}_2|| +\mathfrak{M}(\mathfrak{L}_1,\mathfrak{L}_2)$$
$$\mathfrak{M}(\mathfrak{L}_1,\mathfrak{L}_2) = max\{|\Psi_{\mathfrak{L}_1}|, |\Psi_{\mathfrak{L}_2}|\} - |\Psi_{{\mathfrak{L}}_1} \cap \Psi_{{\mathfrak{L}}_2} |$$
$\Psi_{{\mathfrak{L}}}$ is the multiset of vertex labels in $\mathfrak{L}$.
\end{Definition}

The authors define the distance between two multisets of star structures. Subsequently, they define the mapping distance between two graphs based on the edit distance between their star representations using the Hungarian algorithm~\cite{Kuhn1955,Munkres1957}.

In \cite{Wang2012b}, the authors proposed an index based on the star subdivision provided in \cite{Zeng2009}. This index is made up of two parts: an index for all distinct star structures from the
given database, and an inverted list below each star structure. The star structures are sorted in alphabetical order. 
Each entry in the inverted lists contains the graph identity and the frequency of the corresponding star structure. All lists are sorted in increasing order of the graph size \cite{Wang2012b}. However, enumerating all the different stars in a large graph database may produce a huge index which is not a practical solution.

\begin{figure}[tbph]
\centering
\includegraphics [scale=0.35]{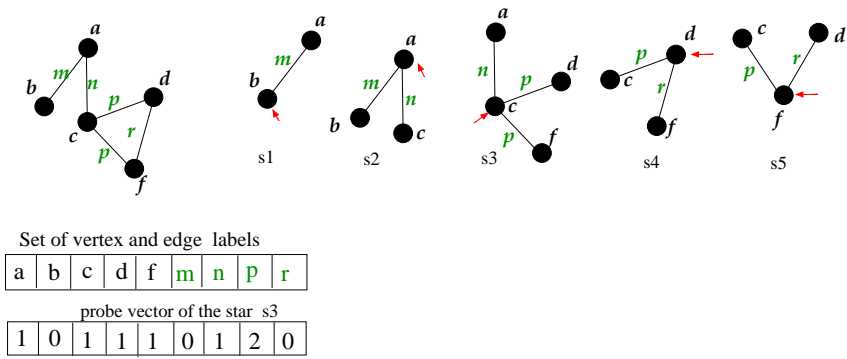}
\caption{\label{fig:Decomposition-Raveaux}A graph and its decomposition into probe vectors.
}
\end{figure}

In \cite{Raveaux2010}, the authors also propose a polynomial time graph matching distance based on subgraph matching using the Hungarian algorithm~\cite{Kuhn1955,Munkres1957}. The subgraphs are also stars but consider edge labels which is not the case with \cite{Zeng2009} and \cite{Riesen2007,Riesen2009}. Each star structure is embedded within a vector of probes. Each probe gives the number of times that a given label appears in the star. An example is  described in Figure \ref{fig:Decomposition-Raveaux}. 

Note that the decomposition into stars in the approaches of  \cite{Riesen2009}, \cite{Zeng2009} and \cite{Raveaux2010} induce more overlappings than the decomposition into BARGs of \cite{Eshera1984a,Eshera1984b} as the number of BARGs is smaller than the number of stars in a graph.

\begin{figure}[tbph]
\centering
\includegraphics [scale=0.35]{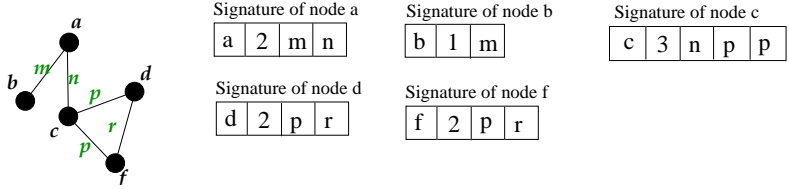}
\caption{\label{fig:Decomposition-Jouili}A graph and its decomposition into node signatures.
}
\end{figure}

Another resembling distance is also defined in \cite{Jouili2009} where a different representation of the star structure is used. In this similarity measure, the star structure is called node signature and is represented by a vector containing the label of the root vertex, its degree, and the set of labels of its incident edges. So, in this representation, the labels of the leaves of the star are not considered in the subgraph as illustrated in Figure \ref{fig:Decomposition-Jouili}. A distance between two node signatures is also defined and the distance between two graphs is then defined as an assignment problem in the matrix containing the distances between nodes signatures of the two compared graphs.

In \cite{Wang2012}, the authors propose to decompose the compared graphs into $k$-Adjacent Tree ($k\_AT$) patterns (like $Q$-Gram decomposition of strings \cite{Sutinen1995}), then use the number of their common $k\_AT$ patterns for edit distance estimation.
The adjacent tree of a vertex $v$ ($AT(v)$) in a graph $G$ is a breadth-first search tree rooted at vertex $v$, the children of each node of $AT(v)$ are sorted by their labels in the graph. The $k$-adjacent tree of
a vertex v ($k\_AT(v)$) in a graph $G$ is the top $k$-level subtree of $AT(v)$ \cite{Wang2012}. This means that the star structure of \cite{Eshera1984a,Eshera1984b} and the related methods is a $1\_AT$.

\begin{figure}[tbph]
\centering
\includegraphics [scale=0.35]{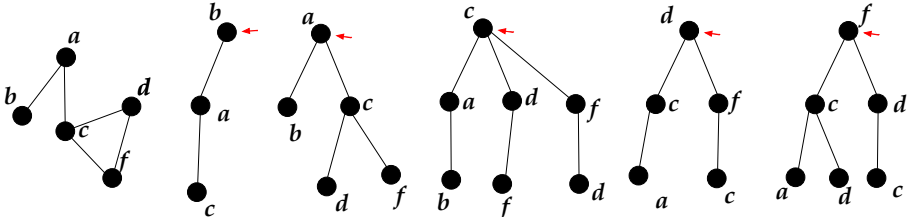}
\caption{\label{fig:Decomposition-KAT}A graph and its $2\_AT$s decomposition.
}
\end{figure}

The set of all $k\_AT$s of a graph $G$ is denoted $k\_ATs(G)$. An example is illustrated in Figure \ref{fig:Decomposition-KAT}. The number of common $k\_ATs$, i.e., $|k\_ATs(G_1)\cap k\_ATs(G_2)|$,  of two graphs is called the matching number of the two graphs and is used to estimate their edit distance
using the following inequality:
\begin{scriptsize}
$$|k\_ATs(G_1)\cap k\_ATs(G_2)|\geq |V(G_1)|-GED(G_1,G_2).2(\Delta(G_1)-1)^{k-1}.$$
\end{scriptsize}
where $GED(G_1,G_2)$ is the edit distance between $G_1$ and $G_2$ and $\Delta(G_1)$ and $\Delta(G_2)$ are the maximum degrees of $G_1$ and $G_2$ respectively with $\Delta(G_2)>1$ and $\Delta(G_2)>1$.
 This estimation has proven to be sufficiently tight but only for sparse graphs \cite{Wang2012}.

To avoid the above cited drawback of tree-based $q$-grams, \cite{Zhao2012} proposes to  use a decomposition into path-based $q$-grams. A path-based $q$-gram in a graph $G$ is a path of length $q$ with no repeated vertex. The edit distance can be estimated with path based $q$-grams with the following inequality:
If $GED(G_1,G_2)\leq\tau$ then $G_1$ and $G_2$ share at least $ max(|Q_{G_1}|-\tau \cdot D_{path}(G_1),|Q_{G_2}|-\tau \cdot D_{path}(G_2))$ path based $q$-grams, where $|Q_{G}|$ is the size of the multiset of path based $q$-grams in $G$ and $D_{path}(G)$ is the number of path based $q$-grams of $Q_G$ affected by an edit operation that occurs on $G$. $D_{path}(G)$ can be computed by:\\
$$D_{path}(G)= max_{u\in V_G}|Q_{G}^u|$$
 where $Q_{G}^u$ denotes the multiset of path $q$-grams that contain vertex $u$.
To find the pairs of graphs that are within an edit distance of $\tau$, the authors propose to use either an inverted index that maps each path $q$-gram to a list of identifiers of
graphs that contain this path $q$-gram or a prefix filter such as those used in string similarity measures \cite{Chaudhuri2006}.

\begin{figure}[tbph]
\centering
\includegraphics [scale=0.35]{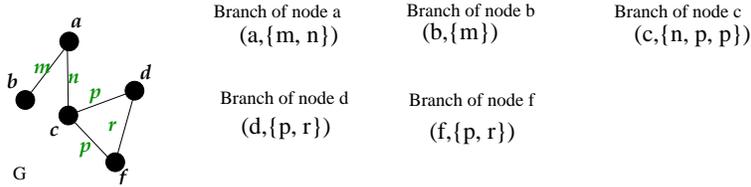}
\caption{\label{fig:Decomposition-Branch}A graph and its branch structures.
}
\end{figure}

In \cite{Zheng2013}, the authors point-out that path-based $q$-grams still induce many overlapping structures. If there are some high-degree vertices, the estimated edit distance of the path-based $q$-grams is not tight. They propose to use a new $q$-gram based structure, called \textit{branch structure}, so that a single edit operation can affect two structures at most allowing a tighter lower bound for edit distance than existing $q$-grams structures.
A branch structure $b$ is a vertex $v$ and the multiset of edge labels incident to $v$. A branch is represented by $b(v)=(l_{v},ES)$, where $l_{v}=L_{V}(v)$ is the label of vertex $v$, and $ES=\lbrace L_{E}(e)\; \vert $ edge $e$ is adjacent to $v$ $\rbrace$ is the multiset of edge labels adjacent to $v$. An example is given in Figure \ref{fig:Decomposition-Branch}. A branch structure is equivalent to the node signature introduced in \cite{Jouili2009}.
Figure~\ref{fig:Decomposition-Branch} shows an example of a graph and its branch structures.
The authors define the edit distance between two branches as in \cite{Zeng2009} and derives the distance between two multisets of branches $B(G_1)$  and $B(G_2)$ as the minimum weighted match in the bipartite graph which vertices represent the branches of $B(G_1)$  and $B(G_2)$ and edges represent transformations between any two branches (from $B(G_1)$  and $B(G_2)$  respectively) weighted with their pairwise branch edit distance. For solving the assignment problem, the authors use the Hungarian algorithm \cite{Kuhn1955}.
The authors also prove that the obtained branch based distance is tighter that the star based distance of \cite{Zeng2009}.

To simplify the processing of a query graph in a large graph database, in \cite{Zheng2013} the authors propose to use an $R$-tree based index where each leaf is the set of branches of a graph of the database. An internal node of the tree is the union of the branches of its children. The query graph is processed by traversing the index starting from the root. For an intermediate node, the branch based distance is computed between the query graph and the set of branches of the internal node. If this distance is greater than a given threshold, the subtree rooted at this internal node can be safely pruned.
However, computing the distance with all the branch set of an internal node is not compatible with large databases and may induce an important overhead.

In \cite{Zhao2013}, the authors propose to use variable-size non-overlapping partitions.  The proposed partitioning is based on the half-edge concept defined as an edge with only one end node and denoted by $(u,.)$.
Based on this concept, the authors introduce the notion of half-edge graph, i.e.,  a graph that contains half-edges, and half-edge subgraph isomorphism defined as follows:
\begin{Definition} \cite{Zhao2013}
A graph $Q=(V_Q,E_Q,f_{V_Q})$ is half-edge subgraph isomorphic to a graph $G=(V_G,E_G,f_{V_G})$, denoted as $Q\sqsubseteq G$, if there exists an injection $h:V_{Q}\rightarrow V_{G}$ such that (1) $\forall u\in V_{Q}, h(u)\in V_{G}$ and $f_{V_Q}(u)=f_{V_G}(f(u))$;(2) $\forall (u,v)\in E_{Q}, (f(u),f(v))\in E_{G}$ and $f_{V_Q}((u,v))=f_{V_G}((f(u),f(v)))$; and (3) $\forall (u,.)\in E_{Q}, (f(u),w)\in E_{G}$ and $f_{V_Q}((u,.))=f_{V_G}((f(u),w)),w\in V_{G}\setminus h(V_{Q})$
\end{Definition}
Based on this, \cite{Zhao2013} develops a partition-based similarity search framework that contains two phases: an indexing phase that can be performed offline and a query processing phase performed for each query. The indexing phase takes as
input a graph database $D$ and an edit distance threshold $\tau$  and constructs an inverted index as follows:
 \begin{itemize}
 \item For each data graph $G\in D$, it first divides $G$ into $\tau +1$ partitions. Figure \ref{fig:Decomposition-halfSG} gives an example of a graph partition into $2$ half edge-subgraphs.
 \begin{figure}[tbph]
\centering
\includegraphics [scale=0.39]{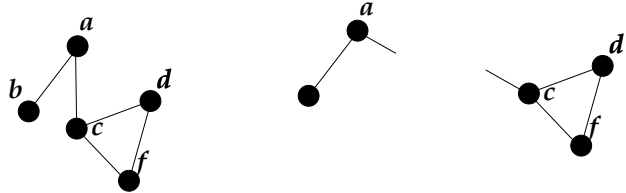}
\caption{\label{fig:Decomposition-halfSG}A half-edge subgraph decomposition.}
\end{figure}
 \item Then, for each partition, it inserts $G$'s identifier into the corresponding postings list of the partition.
 \end{itemize}
 The processing of a query $q$, starts by probing the inverted index for candidate generation. For each partition $p$ in the inverted list, it tests whether $p$ is contained by the query. If so, the graphs in the postings list of $p$ are filtered based on their size and their labels. If the filtering produces a result within $\tau$, the graph is produced as a candidate for the query. Finally, candidates are further examined with a classic graph edit distance algorithm.
The main problem of this approach is related to the partitioning algorithm. In fact, such partitioning is not unique for a given graph. Furthermore, the index is not practical for large graphs.

\begin{figure}[tbph]
\centering
\includegraphics [scale=0.36]{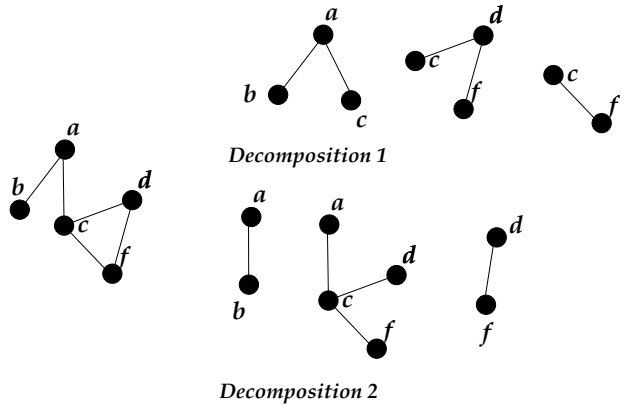}
\caption{\label{fig:Decomposition-STWig}A graph and two of its possible decomposition into STwigs. 
}
\end{figure}

In \cite{Sun2012}, the authors propose a graph decomposition into $STwigs$. An  $STwig$ is a two level tree structure, $q=(r,L)$, where $r$ is the label of the root node and $L$ is the set of labels of its child nodes. Contrarily to the star structure used in \cite{Zeng2009} and \cite{Raveaux2010}, $STwigs$ do not overlap (regarding edges), they are edge disjoint stars as illustrated in Figure \ref{fig:Decomposition-STWig}.

Clearly, such decomposition is not unique and different decompositions of the same query incur different query processing cost.
 So, \cite{Sun2012} proposes a query decomposition that minimizes the number of obtained $STwigs$. The authors proved that the minimum $STwig$ cover problem is polynomial equivalent to the minimum vertex cover problem. Consequently, they construct an $STwig$ cover from a vertex cover in polynomial steps using an existing $2$-approximate algorithm \cite{Cormen2001} for the vertex
cover problem. Given a query graph $q$, \cite{Sun2012} first decomposes $q$  into a set of $STwigs$, then it uses exploration to find matches to each
$STwig$. Exploration at this step avoids indexing on $STwig$s which is not feasible for billion node graphs. Finally, the approach joins the results to find the final solution.
The authors also modified the $2$-approximate algorithm for the $STwig$ cover to incur an $STwig$ order that optimizes the number of joins. In fact, it seems that given a set of $STwigs$ produced by the decomposition
step, an optimized order is the one that ensures that the root node of each $STwig$ is a leaf node of at least one of the already processed $STwigs$.

\begin{figure}[tbph]
\centering
\includegraphics [scale=0.40]{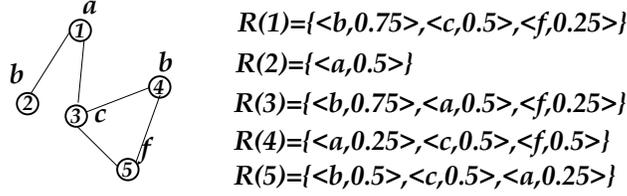}
\caption{\label{fig:Decomposition-NV}A graph and its neighborhood vectors ($h=2$ and $\alpha=0.5$, Vertices are numbered to distinguish them).}
\end{figure}

In \cite{Khan2011}, the authors propose a tree $q$-gram like decomposition embedded in a vector representation. In this approach, each partition rooted at node $u$ encompasses the $h$-hop neighbors of $u$, i.e., the set of nodes $v$ whose distance from $u$ is less than or equal to $h$. The partition is encoded within a multidimensional vector, called \textit{neighborhood vector} and denoted $R(u)$ for node $u$ with $R(u)=\lbrace\langle l,A(u,l)\rangle\rbrace$, where $l$ is a label presents in the neighborhood of $u$ and $A(u,l)$ represents the strength of $l$ in the neighborhood of node $u$ and is obtained by:
\begin{equation}
A(u,l)=\sum_{i=1}^{h}\alpha^{i}\sum_{d(u,v)=i}I(l\in L(v))
\end{equation}
In this formula, $L(v)$ is the label set of node $v$, $I(l\in L(v))$ is an indicator function which takes the value $1$ when $l$ is in the label set of $v$ and $0$ otherwise. $d(u,v)$ is the distance between $u$ and $v$. $\alpha$ is a constant called the propagation
factor that takes value between $0$ and $1$. Figure~\ref{fig:Decomposition-NV} gives an example of a graph and the neighborhood vectors associated to each of its vertices with $h=2$ and $\alpha=0.5$.
The similarity between two neighborhood vectors $R(u)$  and $R(v)$ of two nodes $u$ and $v$ respectively is computed by the following cost function:
\begin{equation}
C(u,v)=\sum_{l\in R(u)} M(A(u,l),A(v,l))
\end{equation}
\begin{equation}
\label{EQ_M}
M(x,y)= \left\{ \begin{array}{ll}
x-y & \text{ if } x > y, \\
0 & \text{ otherwise}.
\end{array} \right.  \\
\end{equation}

Using neighborhood vectors, the authors propose an algorithm that finds all the embeddings of a query graph $Q$ in a target graph $G$ as follows:
\begin{enumerate}
\item compute the neighborhood vectors $R_G(u)$ and $R_Q(v)$ for all nodes $u \in V(G)$, $v\in V(Q)$,
\item for each node pair $u \in V(G)$, $v\in V(Q)$ s.t. $L(v)\subseteq L(u)$,  calculate the node matching
cost, $cost(u,v)$ as the difference of their neighborhood vectors, $cost(u,v) = \sum_{l\in R(v)} M(A_Q(v,l),A_G(u,l))$. Obtaining for each $v\in V(Q)$ a list $List(v)$ of possible matching nodes such that $List(v)= \{u \in V(G), cost(u, v)\leq \varepsilon \}$ where $\varepsilon$ is a similarity threshold. To speed up the computation of $List(v)$ for all $v \in V_Q$, two kinds of indexes can be constructed offline for $G$:
\begin{itemize}
\item a label-based index with a hash table corresponding to each label of $G$. This index is efficient if the labels are node selective.
\item structure-based index  which is built on the neighborhood vectors.
\end{itemize}
\item use dynamic programming to find the embeddings of $Q$ in $V$ from the final list of matched nodes for each node $v \in V_Q$.
\end{enumerate}

 In \cite{Khan2013}, the authors extend the approach proposed in  \cite{Khan2011} with an inference algorithm that iteratively boosts the score of more promising candidate nodes, considering both label and structural similarity. This approach, called $NeMa$ is based on the neighborhood vector introduced in \cite{Khan2011} with the slight difference that the neighborhood vector in $NeMa$ gives more importance to the distance than to the labels. The authors motivate this by two remarks from real applications:  $(a)$  if two nodes are close in a query graph, the corresponding nodes in the
result graph must also be close. However, $(b)$ there may be some differences in labels of the matched nodes due to noises and heterogeneity in data.\\
The neighborhood of a node $u$ in a graph $G$ is given by $R_G(u)=\{<u',P_G(u,u')>\}$, where $u'$ is a node within $h$-hops of $u$, and $P_G(u,u')$ denotes the proximity of $u'$ from $u$ in $G$.
\begin{equation}
P_G(u,u')= \left\{ \begin{array}{ll}
\alpha^{d(u,u')} & \text{ if } d(u,u')\leq h, \\
0 & \text{ otherwise}.
\end{array} \right.  \\
\end{equation}
Where $d(u, u')$ is the distance between $u$ and $u'$ . The propagation factor $\alpha$ is a parameter between $0$ and $1$; and $h > 0$ is the hop number delimiting the neighborhood.
Given a matching function $\phi$, the matching cost of the neighborhood vectors of two nodes $v$ and $u=\phi(v)$ is given by:
\begin{equation}
N_{\phi}(u,v)=\frac{\sum_{v'\in N(v)} M(P_Q(v,v'),P_G(u,\phi(v')))}{\sum_{v'\in N(v)}{P_Q(v,v')}}
\end{equation}
$M$ is defined by  Equation \ref{EQ_M}.

The global cost $C(\phi)$ of the matching function $\phi$ between the query graph $Q$ and the target graph $G$ is given by:
\begin{equation}
C(\phi)=\sum_{v\in V_{Q}} F_{\phi}(v,\phi(v))
\end{equation}
where, $F_{\phi}(v,\phi(v))$ is the individual node matching cost between $v$ and $u$ defined as a linear combination of the label difference function and the neighborhood matching cost function via a parameter $0 < \lambda < 1$, whose
optimal value is set empirically.
\begin{equation}
F_{\phi}(v,\phi(v)) = \lambda \Delta_L (L_Q(v),L_G(\phi(v)) + (1-\lambda) N_{\phi}(v, \phi(v))
\end{equation}

The label difference function $\Delta_L$ between two node labels is defined by the Jaccard similarity.

To find a matching function $\phi$ that minimizes $C(\phi)$, \cite{Khan2013} uses a heuristic based on the max-sum inference problem in graphical models \cite{Pearl1982}.

\subsection{State Space Exploring Approaches}
In these classes of methods, we find mainly 
graph pattern matching approaches where a number of candidates vertices, subgraphs or regions are explored in a large data graph to find the different embeddings of some query graph or the subgraphs that match a given graph pattern.

In \cite{Han2013}, the authors propose a solution, called $TURBO_{ISO}$, to robustly compute subgraph isomorphism with two mechanisms: a tree rewriting of the query graph and candidate region exploration. A candidate region for a query graph $Q$ is a subgraph of the data graph $G$ which may contain embeddings of the query graph. So, performing subgraph isomorphism search on all candidate regions will ensure that all embeddings can be obtained. However, minimizing the number of candidate regions and the size of each region is obviously important for faster matching.
In order to minimize the size of each candidate region, the authors propose to :
\begin{enumerate}
\item rewrite the query $Q$  into an equivalent NEC (Neighborhood Equivalence Class) tree $Q'$. In $Q'$ each set of vertices that have the same label and the same set of adjacent query vertices are merged into one NEC vertex. So, a NEC vertex is a compressed form of a set of vertices. Consequently, using $Q'$ instead of $Q$,  will accelerate the candidate region exploration process, since the number of vertices is smaller.
 \item construct candidate regions for the query $Q$ in the data graph $G$ by constructing for each region a BFS search tree $T_{G}$ from the root node $u'_{s}$ of the NEC tree $Q'$ so that each leaf is on the shortest path from $u'_{s}$. Then, for the start vertex $v_{s}$ of each target candidate region, identify candidate data vertices for each query vertex by simply performing depth-first search using $T_{G}$ and starting from $v_{s}$.
\end{enumerate}
Minimizing the number of regions comes through a careful choice of the root of the NEC tree. 
For this, $TURBO_{ISO}$  ranks every query vertex $u$  by $Rank(u) = \frac{freq(G,L(u))}{deg(u)}$ , where $freq(G, l)$ is the number of data vertices in $G$ that have label $l$, and $deg(u)$ means the degree of $u$. This ranking function favours lower frequencies and higher degrees which will minimize the number of regions.

When exploring candidate regions, $TURBO_{ISO}$ also minimizes the number of enumerated partial solutions by ordering the NEC vertices by increasing sizes. Thus, paths involving fewer vertices are explored first, the space is pruned rapidly if no isomorphism is possible.

\cite{Fan2010} introduces bounded simulation, an extension of graph simulation intended to deal with graph queries expressed with graph patterns. In this case, all graphs are directed and a pattern graph is defined  as follows: 
\begin{Definition} \label{Def-Pattern}\cite{Fan2010}
A pattern graph is defined as $P=(V_P,E_p,f_{V_P},f_{E_P})$, where
\begin{enumerate}
 \item $V_P$ and $E_P$ are the set of nodes and the set of directed edges, respectively, as defined for data graphs;
 \item $f_{V_P}$ is a function defined on $V_P$ such that for each node $u$,$f_V(u)$ is the predicate of $u$, defined as a conjunction of atomic formulas of the form $A\; op\; a$; here $A$ denotes an attribute, $a$ is a constant, and $op$ is a comparison operator $<,\leq, >, \geq,=,\neq$;
 \item $f_{E_P}$ is a function defined on $E_p$ such that for each edge $(u,u')\in E_P$, $f_{E_P}((u,u'))$ is either a positive integer $k$ or a symbol $*$.
\end{enumerate}
\end{Definition}

Intuitively, the predicate $f_V(u)$ of a node $u$ specifies a search condition and may induce several possible label values. The integer $f_{E_P}((u,u'))$ of an edge $(u,u')$ means that the edge $(u,u')$ can be matched to a path of length at most $f_{E_P}(u,u')$. 
A simple graph query corresponds to a graph pattern where $f_V(u)$ is simply the label of $u$ and $f_{E_P}((u,u'))=1$.
In bounded simulation, the term "bounded" relates to the bound piggybacked by each edge in the pattern. This bound is the maximum length of a path in the data graph that matches the edge of the pattern. Bounded simulation is defined as follows:
\begin{Definition}\cite{Fan2010}
 A data graph $G = (V,E,f_{A})$ matches the pattern query $Q = (V_{Q},E_{Q},f_{V_Q},f_{E_Q})$ via bounded simulation, denoted by $Q\trianglelefteq G$, if there exists a binary relation $S\subseteq V_{Q}\times V$ such that:
\begin{itemize}
\item for each $u\in V_{Q}$, there exists $v\in V$ such that $(u,v)\in S$;
\item for each $(u,v)\in S$, (a) the attributes $f_{A}(v)$ of $v$ satisfies the predicate $f_{V_Q}(u)$ of $u$; 
    and (b) for each edge $(u, u')$ in $E_{V_Q}$, there exists a non empty path $\rho=v\diagup...\diagup v'$ in $G$ such that $(u',v')\in S$, and $len(\rho)\leq k$ if $f_{V_Q}(u,u')$ is a constant $k$.
\end{itemize}

\end{Definition}

\begin{figure}[h]
\centering
\includegraphics [scale=0.45]{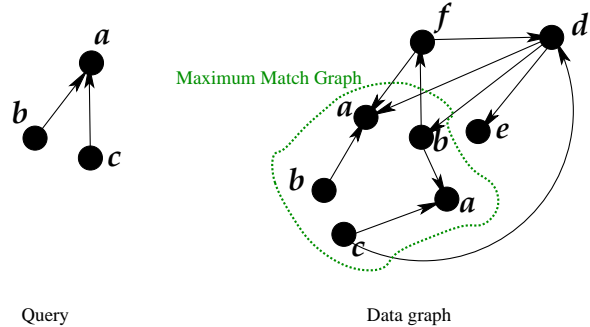}
\caption{\label{fig:Bounded-Sim}Maximum match graph for bounded simulation.}
\end{figure}

In this paper, the authors also introduce the concept of \textit{maximum match graph} to represent the union of all matches of a query in a data graph.
This means that bounded simulation will search for a unique result graph that encompasses all the subgraphs that match the query pattern as illustrated in Figure \ref{fig:Bounded-Sim}. 

Then, they  propose an algorithm for \emph{incremental matching} that avoids the cost related to re-computing the result graph when the graph data is modified. This ensures the scalability of the approach to large graphs.

In \cite{Ma2011}, the authors focused on reducing the number of matches returned by graph simulation and bounded simulation, the extension of graph simulation proposed in \cite{Fan2010}. This is achieved by enforcing two conditions:
\begin{enumerate}
\item Duality which corrects the behavior of graph simulation concerning topology preservation of the query. In fact, as shown in Figure \ref{Fig-iso-sim}, graph simulation may return a disconnected subgraph for a connected graph query which augments the number of matches. To avoid this, \cite{Ma2011} proposes \textit{dual simulation} defined as follows:
   \begin{Definition}\cite{Ma2011}
 A data graph $G = (V,E,f_{A})$ matches the pattern $Q = (V_{Q},E_{Q},f_{v},f_{e})$ via dual simulation, denoted by $Q\prec_{sim}^D G$, if $Q\prec G$ with a binary match relation $S_D \subseteq V_Q\times V$, and for each pair $(u,v)\in S_D$ and each edge $(u_2, u)\in E_Q$, there exists an edge $(v_2,v)\in E$ with $(u_2,v_2)\in S_D$.
\end{Definition}
Thus, dual simulation requires that two related nodes have the same edges and by the way avoids to simulate a connected graph with a disconnected one. Accordingly, in the example of Figure~\ref{Fig-iso-sim} only subgraph $G_1$ is returned as the result graph match.
\item Locality which reduces the diameter of the returned subgraph of bounded simulation.  
    In fact, bounded simulation returns a maximum match that encompasses all the matches of the query. This maximum match is unique but may be a too large graph. Locality is enforced by requiring matches to be within a ball of radius equal to the diameter of the query. A ball is defined as follows:
    \begin{Definition}
     For a node $v$ in a graph $G$ and a non-negative integer $r$, the ball with center $v$ and radius r is a subgraph of $G$, denoted by $\hat{G}[v,r]$, such that (1) for all nodes $v'\in \hat{G}[v,r]$, the shortest distance $dist(v,v')\prec r$, and (2) it has exactly the edges that appear in $G$ over the same node set.
     \end{Definition}
\end{enumerate}

\begin{figure}[tbph]
\centering
\includegraphics [scale=0.4]{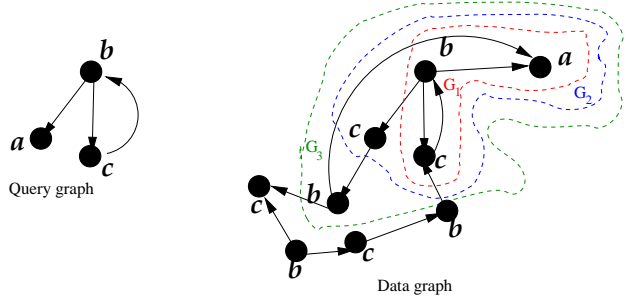}
\caption{\label{fig:Strong-Sim}Bounded simulation Vs Dual and Strong simulation.}
\end{figure}

\begin{Definition}\cite{Ma2011}
 A data graph $G = (V,E,f_{A})$ matches the query pattern $Q = (V_{Q},E_{Q},f_{v},f_{e})$ via strong simulation, denoted by $Q\prec_{sim}^{S} G$, if there exist a vertex $v\in V$ and a connected subgraph $G_{s}$ of $G$ such that:
\begin{itemize}
\item $Q\prec_{sim}^D G_{S}$ with the maximum match relation $S$;
\item $G_{s}$ is exactly the match graph of $Q$ with $S$, and
\item $G_{s}$ is contained in the ball $\hat{G}_{D}[v,d_{Q}]$ of center $v$ and radius $d_{Q}$ the diameter of $Q$. %
\end{itemize}

\end{Definition}
Figure \ref{fig:Strong-Sim} illustrates an example adapted from \cite{Ma2011}. In this example, with simulation and bounded simulation the query graph matches all the data graph. However, dual simulation returns $G_3$ while strong simulation returns $G_2$. Note that the diameter of the query graph is $2$. Subgraph isomorphism returns $G_1$.

The authors show that strong simulation has the same complexity than simulation and bounded simulation while preserving graph topology. They propose a cubic-time algorithm that returns the set of subgraphs of a data graph  that matches by strong simulation a graph query. The algorithm inspects the balls of radius equal to the query diameter and centred at each node of the data graph.

In \cite{Fard2013}, the authors propose \textit{strict simulation} to further improve graph simulation and adapt its computation within a vertex-centric Bulk Synchronous Parallel (BSP) programming model \cite{Valiant1990} used by several graph processing frameworks such as Pregel \cite{Malewicz2009}. They introduce an extra step in the algorithm of strong simulation proposed in \cite{Fan2010}. Strict simulation reduces the size, i.e., the number of nodes, of the ball inspected by strong simulation. For this, the idea is to first compute the match for dual simulation before inspecting the balls. So, the balls are computed on the result of dual simulation and are consequently much smaller than those computed by strong simulation. Formally, strict simulation is defined as follows:
\begin{Definition}\cite{Fard2013}
 A data graph $G = (V,E,f_{A})$ matches the query pattern $Q = (V_{Q},E_{Q},f_{v},f_{e})$ via strict simulation, denoted by $Q\prec_{sim}^{\sum} G$, if there exists a vertex $v\in V$ such that:
 \begin{itemize}
\item $v\in V_{D}$ where $G_{D}(V_{D},E_{D},l_{D})$ is the result match graph with respect to $Q\prec_{sim}^{D} G$;
\item $Q\prec_{sim}^{D} \hat{G}_{D}[v,d_{Q}]$ where $\hat{G}_{D}[v,d_{Q}]$ is a ball extracted from $G_{D}$; and
\item $v$ is a member of the maximum match graph.
\end{itemize}
\end{Definition}

In the example of Figure \ref{fig:Strong-Sim}, strong simulation will first compute the match graph for dual simulation, i.e., subgraph $G_3$, and then  begin inspecting the balls.

\cite{Fard2013} also proposes distributed algorithms to compute simulation, bounded simulation, strong simulation and strict simulation.


Similarly to strict simulation, \cite{Fard2014} introduces \textit{tight simulation} that improves strict simulation and approaches subgraph isomorphism. Tight simulation focuses on reducing the number of the balls inspected by strict simulation.  To do so, the authors propose to select a single vertex $u$ of the pattern $Q$ and to use it as a candidate match to the center of a potential ball in the data graph. $u$ is chosen to be the vertex of minimum eccentricity, i.e., it is a center of $Q$, which has the highest ratio of degree to label frequency (in $Q$). This allows to reduce the radius of the balls and also their number. So, tight simulation is defined as follows:
\begin{Definition}\cite{Fard2014}
 A data graph $G = (V,E,f_{A})$ matches the query pattern $Q = (V_{Q},E_{Q},f_{v},f_{e})$ via tight simulation, denoted by $Q\prec_{sim}^{T} G$, if there  are vertices $u\in Q$ and $u'\in G$ such that
 \begin{itemize}
\item $u$ is a center of $Q$ with highest defined selectivity;
\item $(u,u')\in R_{D}$ where $R_{D}$ is dual relation set between $Q$ and $G$;
\item $Q\prec_{sim}^{D} \hat{G}_{D}[u',r_{Q}]$ where $\hat{G}_{D}[u',r_{q}]$ is a ball extracted from $G_{D}(V_{D},E_{D},l_{D})$ which is the result match graph with respect to $Q\prec_{sim}^{D} G$, and $r_{Q}$ is the radius of $Q$, and
\item $u'$ is a member of the resulting maximum match graph.

 \end{itemize}

\end{Definition}
In the example of Figure \ref{fig:Strong-Sim}, the node having label $b$ is a center of the query graph and will be used to extract the balls in the result of dual simulation, i.e., the match graph $G_3$.
%
The authors show that tight simulation has better results than strong simulation and strict simulation.

\subsection{Summary-based approach}

Graph summarizing/compression offers interesting perspectives for large graph storage and processing. A graph summarizing method that retains an "acceptable amount" of the graph properties may be used as a preprocessing step to several graph algorithms. The idea here is not to reduce the size of a huge graph just to minimize its storage requirement and to decompress the graph to process it. Rather, the aim is to obtain a compressed representation of the graph that can be used, instead of the original graph, by the processing algorithms, i.e., analysis, mining, comparison, querying, etc.
In this vein, \cite{Chen2009} proposes an algorithm that finds all frequent subgraphs in a database of large graphs where the database graphs are summarized. Summarizing is achieved by grouping the nodes that have the same label into  supernodes 
as follows:
\begin{Definition}(Summarized Graph) \cite{Chen2009} \label{Def-Sum}. Given a labeled graph $G$ such that its vertices $V(G)$ are partitioned into groups, i.e., $V(G)=V_1(G), V_2(G),  \cdots, V_k(G)$, such that:
(1) $V_i(G) \cap V_j(G)=\phi, 1\leq i\neq j \leq k$ \\
(2) all vertices in $V_i(G), 1\leq i \leq k$, have the same labels.\\
We can summarize $G$ into a compressed version $comp(G)$ where:\\
(1) $comp(G)$ has exactly $k$ nodes $v_1,v_2, \cdots, v_k$ that correspond to each of the groups of $V(G)$ (i.e., $V_i(G)\mapsto v_i$). The label of $v_i$ is set to be the same as those vertices in $V_i(G)$, and \\
(2) an edge $(v_i,v_j)$ with label $l$ exists in $comp(G)$ if and only if there is an
edge $(u,u')$ with label $l$ between some vertex $u\in V_i(G)$ and some other vertex $u'\in V_j(G)$.
\end{Definition}
The obtained summarized graphs may then be mined for frequent patterns using any existing algorithm. To ensure that all patterns are found, the authors do not systematically summarize all the graphs of the database, rather they proceed with several iterations each of which consists of two steps:
\begin{itemize}
\item Step 1: For each $G_i$ in a graph database D, randomly partition its vertex set $V(G_i)$.
\item Step 2: Execute a pattern mining algorithm of the resulting summarized database.
\item Step 3: Compute the support of each resulting pattern in the original database, i.e., the number of graphs that contain the pattern. Discard the pattern if its support is lower than a predefined threshold.
The number of iteration is controlled by the probability of missing a frequent pattern.

\end{itemize}

In \cite{Fan2012b}, the authors observe that users typically adopt a class $Q$ of queries when querying a data graphs $G$. They  propose a graph compression preserving queries of Q. This means that each query in $Q$ returns the same result when applied to $G$ and when applied to the compression of $G$. They define the compression functions for two kind of graph queries: reachability queries and pattern queries. Roughly speaking, for reachability queries which aims to define if a node is reachable from another, the compression function groups the nodes that have the same ancestors and the same descendants. For pattern queries, the compression function is equivalent to the one given by Definition \ref{Def-Sum}.



\begin{figure*}[ht]
\centering
\begin{subfigure}[b]{0.4\textwidth}
                \centering
               \includegraphics[width=5cm]{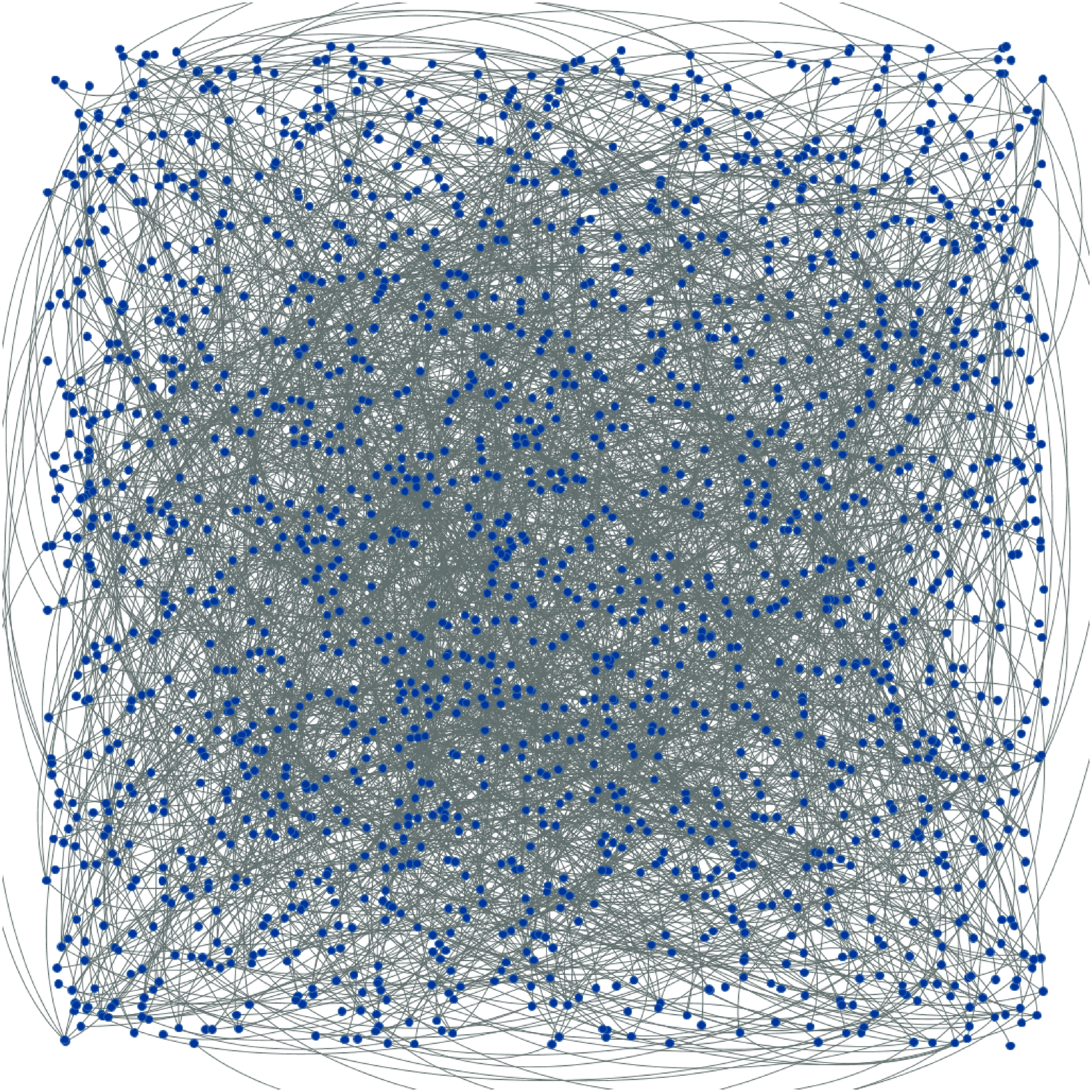}
                \caption{A protein graph of 1818 nodes and 1833 edges from the RI database \cite{Bonnici2013}.}
                \label{fig:graph}
        \end{subfigure}
        \quad
        ~ 
        \begin{subfigure}[b]{0.4\textwidth}
                \centering
              \includegraphics[width=5cm]{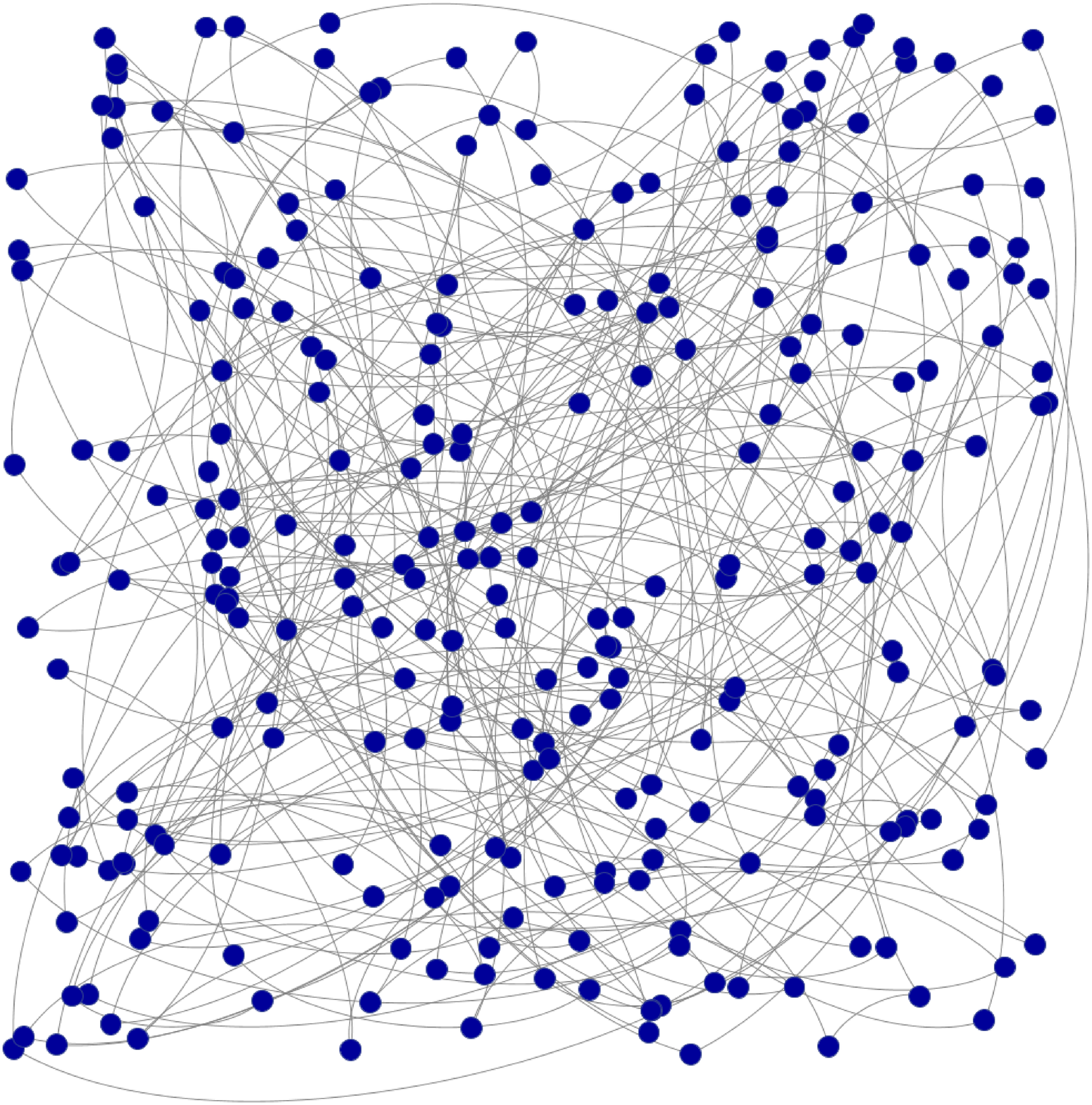}
                \caption{The corresponding prime graph having 271 nodes and 321 edges.}
                \label{fig:prime}
        \end{subfigure}

\caption{Example of a graph and its prime graph.}
\label{fig:graphAndPrime}
\end{figure*}

In~\cite{Lagraa2014}, the authors propose a new solution for the comparison of large graphs. Their approach relies on a compact encoding of graphs called \textit{prime graphs}. Prime graphs are smaller and simpler than the original ones but they retain the structure and properties of the encoded graphs. An example of a graph and its prime is given in Figure~\ref{fig:graphAndPrime}. In~\cite{Lagraa2014}, the authors propose to approximate the
similarity between two graphs by comparing the corresponding prime graphs. Their proposed approach involves the following steps:
\begin{itemize}
\item Building the prime graph of the compared graphs. Prime graphs are obtained by  modular decomposition of the original graphs. 
    Modular decomposition is one of the most known graph decompositions~\cite{Habib2010}. It was introduced by Gallai~\cite{Gallai1967} to solve optimization problems. Modular decomposition generates a representation of a graph that highlights groups of vertices that have the same neighbors outside the group. These subsets of vertices are called \textit{modules}. The prime graph correspond to the graph obtained by compressing all the modules recursively.
\item Partitioning the compared prime graphs into stars of modules as in ~\cite{Zeng2009}.
\item Computing the distance between two prime graphs  based on the distance of each pair of the stars of modules. Given a query prime graph $PG_1$ and a target prime graph $PG_2$, the nodes of $PG_1$ are mapped to the nodes of $PG_2$ using the Hungarian algorithm by defining a cost matrix that records for each star of modules from $PG_1$ the edit operations that are needed to transform it to each star of modules of $PG_2$.

\item Solving the assignment problem by using the Hungarian algorithm~\cite{Kuhn1955} to obtain the minimum distance.

\end{itemize}


\section{Discussion}
\label{sec:Discussion}

Tables~\ref{Table1},~\ref{Table2},~\ref{Table3} summarize all the presented approaches within the three categories: partition-based approaches, search-space exploring approaches  and  summary-based approaches, respectively.

The tables summarize these approaches according to the following facets:
\begin{itemize}
    \item Graphs: the type of graphs on which the graph comparisons are performed: directed/undirected graph, labeled/unlabeled edges.
    \item Decomposition unit: the type of graph partitioning given by the name of the subgraph structure.
    \item Comparison concept: the type of similarity used for graph comparison.
    \item Application: describes the application area of the approach.
    \item Program: the type of program, it can be sequential or parallel.
    \item Size of the query: describes the range of the size of the graph query used for matching.
    \item Size of data graph: describes the range of the size of the data graph used for matching. The size here is given in terms of the number of nodes and edges in the graph. It can be thousand (k), million (M) or billion.
    \item Time complexity of the approach when computed.

\end{itemize}

\centering
\begin{sidewaystable*}

\centering
\caption{Summary of partition-based approaches}
\label{Table1}

\begin{scriptsize}

 \begin{center}
\begin{tabular}{|c c c c c c c c c|}
\hline
\textbf{Approach} & \textbf{Graphs} & \textbf{Decomposition} & \textbf{Comparison }&\textbf{Application} & \textbf{Size of}  & \textbf{Size of}    & \textbf{Program} & \textbf{Time}  \\ 
         &      &  \textbf{unit}   &  \textbf{concept}  &            &\textbf{the query} & \textbf{data graph} &               & \textbf{Complexity} \\
\hline
\hline
\cite{Eshera1984a,Eshera1984b} & directed                & BARG & Edit          &Image        & No experiment  & No experiment & Sequential  & $O(|V_{G}|^2|V_{Q}|^2(|V_{G}|+|V_{Q}|))$  \\
                               &  labeled         &      &   distance    &processing  & &  & &  \\
                                &  edges         &      &       & & &  & &  \\
\hline
\cite{Shapiro1985} & directed                & Relational & Number of          &Image         &  No experiment & No experiment  & Sequential  & Not computed  \\
                   & unlabeled          &  Description   &  common RDs     &processing  &     & & &  \\
&  edges         &     &       &  &     & & &  \\
\hline
\cite{Raveaux2010} &undirected      & Star & Edit     & Image       & 4-12 nodes   & 4-12 nodes  & Sequential  & $O(V_{G}^{3}$)  \\
                   &labeled    &      & distance &  processing  &   3-11  edges& 3-11  edges & &  \\
                   &  edges              &      & Probing  &               &             &             & &  \\
\hline
\cite{Zeng2009} & undirected         & Star  &Edit     & Chemistry &  5-65 nodes & 1 - 80 nodes  & Sequential & $O(V_{G}^{3})$  \\
                & unlabeled    &       & distance&  Networks & $\simeq 30$ edges &   -       &        & \\
& edges    &       & &  &   &   -       &        & \\
\hline
\cite{Riesen2007,Riesen2009}   & undirected                  & Star & Edit         &Image        & -  &8-126 nodes  & Sequential  &  Not computed 
\\
                               &  labeled          &      &  distance     &processing  & - &9-328 edges & &\\
                              & edges         &      &       &  && & &\\
\hline
\cite{Jouili2009} & undirected & Signature & Edit         &    Retrieving     & -   & 9-417 nodes  & Sequential  & Not computed  \\
                 &  labeled         &     &  distance     & Image  & - & 9-112 edges & &  \\
& edges         &      &       &  && & &\\
\hline
\cite{Zhao2013} & undirected& Half-edge & Edit & Chemistry & - & 40 - 100k nodes & Sequential  &   \\
                & unlabeled  & subgraph  &distance & Networks &- & - &  &   \\
                & edges         &      &       &  && & &\\
\hline
\cite{Zhao2012} &undirected& path-based&Edit  & Chemistry  & - & 40 - 126 nodes  & Sequential &   $O(\tau(\vert V_{G}\vert+\vert V_{Q}\vert)log\vert V_{Q}\vert)$ \\
                &unlabeled  & $q$-gram & distance & Networks & -  & - &  &   \\
                & edges         &      &       &  && & &\\

\hline
\cite{Wang2012} &undirected        & $k\_AT$ & Edit  & Chemistry & - & 40 - 100k nodes & Sequential &  Not computed \\
                &unlabeled   &       &  distance            & Networks  & - &    -       &        &   \\

& edges         &      &       &  && & &\\
\hline

\cite{Sun2012} & undirected & STwig & Subgraph & Web       & 3-10 nodes & 80 - 4096K nodes  & Parallel & $O(|q|^{3})$  \\
                & unlabeled &     &  Matching & Networks &   10-20  edges &   -   &  & \\
                & edges         &      &       &  && & &\\
\hline

\cite{Khan2011} & undirected & Neighborhood       & Edit    & Web       & 8-12 nodes & 172k-100000k nodes & Sequential  & $O(|V_{G}|.d^h)$  \\
                & unlabeled & vector & distance & Networks & - & 579k-213000k edges &  &  \\
                & edges         &      &       &  && & &\\
\hline

\cite{Khan2013} &undirected& Neighborhood & Edit & Web  & 3-7  nodes&  2M - 12M nodes & Sequential  & $O(\vert V_{Q}\vert.\vert V\vert+I.\vert V_{Q}\vert.m_{Q}^{2}.d_{Q})$ \\
                          & unlabeled &vector & distance & Networks & - & 11M - 20M edges &  &  \\
                          & edges         &      &       &  && & &\\
\hline

\cite{Zheng2013} & undirected& Branch  & Edit &  Biology & 40k - 100k nodes & 40k - 100k nodes & Sequential  & Not computed \\
               & labeled & structure & distance &   & - & - &  &  \\
               & edges         &      &       &  && & &\\
\hline

\end{tabular}
 \end{center}
\begin{flushleft}
\hspace*{1cm}
$\tau$: graph edit distance threshold.\\
\hspace*{1cm}
\textit{h}: hops. \textit{d}: the average degree of each node.\\
\hspace*{1cm}
$d_{Q}$: the maximum number of $h-hop$ neighbors of each query node.\\
\hspace*{1cm}
$m_{Q}$ maximum number of candidates per query node.
\end{flushleft}
\end{scriptsize}

\end{sidewaystable*}

\begin{sidewaystable}[h]
\caption{Summary of search-space exploring approaches}
\label{Table2}
\begin{scriptsize}
    \centering
\begin{tabular}{|c c c c c c c c|}
\hline
\textbf{Approach} & \textbf{Graphs} & \textbf{Comparison Concept} &\textbf{Application} & \textbf{Size of}  & \textbf{Size of}    & \textbf{Program} & \textbf{Time}  \\ 
         &      &               &            &\textbf{the query} & \textbf{data graph} &               & \textbf{Complexity} \\
\hline
\hline

\cite{Han2013} & undirected& Subgraph & Biology  & 2-15  nodes & 0.5M-4M  nodes  & Sequential  & $O(|V_{G}^2|)$ \\
               & unlabeled& isomorphism & & 1-10  edges & 32M  edges &  &  \\

\hline

\cite{Fan2010} & directed &Bounded & Web  & 4-10  nodes & 1k-20k  nodes  & Sequential  & $O(|V_{G}||E_{G}|+$ \\
               & labeled edges & Simulation &  & - & 19k-58k  edges &  & $|E_{Q}||V_{G}|^2+|V_{Q}||V_{G}|)$ \\
\hline

\cite{Ma2012} & directed &Strong  & Web  & 3-15  nodes & millions of nodes & Sequential  & Not computed \\
              & unlabeled &simulation &  & - & billions of edges &  & \\
\hline

\cite{Fard2013} &directed&Strict  & Social & 10-20  nodes &millions of nodes & Parallel & Not computed \\ 
 &unlabeled edge &simulation &  Networks & - & billions of edges &  & \\
\hline

\cite{Fard2014} & directed&Tight  & Social & 5-100  nodes &millions of nodes & Parallel & $O(\vert V_{q}^{3}\vert)$ \\ 
 &unlabled &simulation &  Networks & - &billions of edges &  & \\
\hline
\end{tabular}
\end{scriptsize}
\end{sidewaystable}

\begin{sidewaystable}[h]
\caption{Existing summary-based approaches}
\label{Table3}
\begin{scriptsize}
 \begin{center}
\begin{tabular}{|c c c c c  c c c|}
\hline
\textbf{Approach} & \textbf{Graphs} & \textbf{Comparison Concept} &\textbf{Application} & \textbf{Size of}  & \textbf{Size of}    & \textbf{Program} & \textbf{Time}  \\ 
         &      &               &            &\textbf{the query} & \textbf{data graph} &               & \textbf{Complexity} \\
\hline
\hline

\cite{Chen2009} & undirected &Subgraph & Program   & Not & 100-20k  nodes  & Sequential  & Not computed \\
               & labeled edges & mining &  data & Necessary & 220k  edges &  &  \\

\hline

\cite{Fan2012b} & directed & Compression & Social   & 3-8 of nodes & 6k-2.4M  nodes  & Sequential  & $O(|V(G)|^2 + $ \\
               & labeled edges & preserving query &  Networks & 3-8  edges & 21k-5M  edges &  & $|V(G)||E(G)|)$ \\

\hline

\cite{Lagraa2014} & undirected & Prime & Biological  & 8-34000   nodes & 9-33k  nodes  & Sequential  & $O(k^3 + $ \\
               & unlabled edges & graph & graphs & & 9-332k  edges &  & $|V(G)| + |E(G)|)$ \\
\hline
\end{tabular}
 \end{center}
$k$ is the number of vertices in the largest prime graph.
\end{scriptsize}
\end{sidewaystable}

%
%
%
%
%
%
%
%
%
%
\justify
Throughout this survey, we can see that various solutions are considered and there is not  a generic algorithm for graph comparison or graph pattern matching that takes into consideration any type of graph (labeled/unlabeled and directed/undirected). Partition based approaches become increasingly used for graph comparison and pattern matching approaches. In fact, these approaches have a good time complexity and are easy to project toward parallel algorithms. The problem of matching in partition-based approaches is simplified by decomposing the graphs to be matched into smaller subgraphs. However, the best graph decomposition technique that should be adopted for computing distance remains an open problem for large graphs even if we note that the majority of partitioning approaches relay on a star decomposition.
Besides, the approaches that use the Hungarian algorithm~\cite{Kuhn1955,Munkres1957} on a large cost matrix such as \cite{Zeng2009} suffer memory problems. Heuristics or other methods that compute the minimum cost while avoiding the construction of the cost matrix are appreciated. Also, a parallel version of the Hungarian algorithm that relies on a partitioning of the matrix storage and computation will scale these approaches to larger graphs.

Furthermore, using partition based approaches in subgraph search is generally associated with joins or indexing methods. Both of them are time consuming and complex tasks especially for large graphs. So, research must focus  on methods to avoid them  or develop them to deal with large graphs.

We can also note that several graph matching techniques have not been investigated in large scale graphs Among these solutions we can cite clustering based methods and polynomial heuristics to the greatest common subgraph. Invariant-based graph comparison \cite{Mckay1981,Xiao08} may also give good results.

To cope with large graphs, one among the solutions is graph compression without loss of information and performing the matching on the compressed graph.
However, it does not exist enough summary-based approaches. Reducing and compressing a graph for graph matching is a very interesting approach. There are two benefits: obtaining more storage space in the hard disk and performing the matching in a compressed and reduced graph without decompression~\cite{Lagraa2014}. In addition, graph compression techniques that retain all the information of the original graphs and that can be used for matching  remain a challenge.

In the majority of approaches, the space complexity of graph matching  has not been investigated. The different  approaches do not deal much about space complexity  and memory consumption  of algorithms which are important performance metrics either  in theory or practice coping with large graphs.

The problem of matching dynamic graphs has not received enough interest in the literature. Currently with social networks and the web, graphs change continuously: new nodes and edges are added or deleted from the graph through time. The problem is then to take into consideration the evolution of dynamic graphs in graph comparison or pattern matching approaches. Apart from the work of \cite{Fan2010} we found little literature on this question.


\balance

\section{Conclusion}
\label{sec:conclusion}

The dominance of graphs as a representation tool in real world applications demand new graph matching techniques, concepts, and languages to match large graph datasets efficiently.
We have presented a review of recent works on graph comparison and graph pattern matching approaches on large graphs, highlighting the different notions, techniques and concepts used for matching and their impact coping with large graphs. We classified the approaches into three categories: partition based approaches, search space exploring approaches and summary-based approaches. Each of them has its advantages and application areas.
Many recent graph comparison and graph pattern matching approaches converge towards partitioning of the compared graphs. The problem is simplified by decomposing the graphs to be matched into smaller subgraphs. However, these approaches are not always possible and there are few algorithms suitable for all kinds of graphs and applications.
Globally and as discussed in the previous  section several problems and area of investigations deserve future research despite the substantial results of current and past investigations.
According to the International Technology Roadmap for Semiconductors (ITRS), as many as 6000 processors are expected on a single system-on-chip by the end of year 2026. Moreover, the memory size will follow the same trends. Thus, parallel graph matching algorithm is needed for the next generation  in order to run quickly the matching processes and exploit efficiently the hardware resources such as the number of processors and memory size. Moreover, due to the huge size of graphs, compressing graphs for matching without decompression remains a challenging issue. Combining parallelism with compressing or partitioning is also very  interesting. Furthermore, dynamic graphs and graphs in streaming applications are not sufficiently addressed in the actual research effort.

\bibliographystyle{abbrv}
\bibliography{MyPubli,Bigraphs,Prime,IntroBib}

\begin{thebibliography}{10}

\bibitem{Aggarwal2010}
C.~C. Aggarwal and H.~Wang.
\newblock {\em Managing and mining Graph data}.
\newblock Springer, 2010.

\bibitem{Ambauen2003}
R.~Ambauen, S.~Fischer, and H.~Bunke.
\newblock Graph edit distance with node splitting and merging and its
  application to diatom identification.
\newblock {\em Graph Based Representations in Pattern Recognition - GBR}, pages
  95--106, 2003.

\bibitem{Basu2006}
M.~Basu and T.~K.~H. BBA.
\newblock {\em Data Complexity in Pattern Recognition}.
\newblock Springer, 2006.

\bibitem{Bonnici2013}
V.~Bonnici, R.~Giugno, A.~Pulvirenti, D.~Shasha, and A.~Ferro.
\newblock A subgraph isomorphism algorithm and its application to biochemical
  data.
\newblock {\em BMC Bioinformatics}, 14(Suppl 7)(S13), 2013.

\bibitem{Borgwardt2005}
K.~Borgwardt and H.-P. Kriegel.
\newblock Shortest-path kernels on graphs.
\newblock In {\em 5th Int. Conference on Data Mining}, pages 74--81, 2005.

\bibitem{Bunke1997b}
H.~Bunke.
\newblock {On a relation between graph edit distance and maximum common
  subgraph}.
\newblock {\em Pattern Recognition Letters}, 18:689--694, 1997.

\bibitem{Bunke1999}
H.~Bunke.
\newblock Error correcting graph matching: On the influence of the underlying
  cost function.
\newblock {\em IEEE Trans. Pattern Anal. Mach. Intell.}, 21(9):917--922, 1999.

\bibitem{Bunke1983}
H.~Bunke and G.~Allerman.
\newblock Inexact graph matching for structural pattern recognition.
\newblock {\em Pattern Recognition Letters-PRL}, 1(4):245--253, 1983.

\bibitem{Bunke1997}
H.~Bunke and B.~T. Messmer.
\newblock {Recent Advances in Graph Matching}.
\newblock {\em International Journal of Pattern Recognition and Artificial
  Intelligence}, 11:169--203, 1997.

\bibitem{Bunke2011}
H.~Bunke and K.~Riesen.
\newblock {Recent advances in graph-based pattern recognition with applications
  in document analysis}.
\newblock {\em Pattern Recognition}, 44:1057--1067, 2011.

\bibitem{Bunke1998}
H.~Bunke and K.~Shearer.
\newblock A graph distance metric based on the maximal common subgraph.
\newblock {\em Pattern Recognition Letters}, 19(3-4):255--259, 1998.

\bibitem{Hancock2001}
M.~Carcassoni and E.~R. Hancock.
\newblock Weighted graph-matching using modal clusters.
\newblock pages 142--151, 2001.

\bibitem{Chaudhuri2006}
S.~Chaudhuri, V.~Ganti, and R.~Kaushik.
\newblock A primitive operator for similarity joins in data cleaning.
\newblock In {\em Proceedings of the 22Nd International Conference on Data
  Engineering}, ICDE '06, pages 5--, Washington, DC, USA, 2006. IEEE Computer
  Society.

\bibitem{Chen2009}
C.~Chen, C.~X. Lin, M.~Fredrikson, M.~Christodorescu, X.~Yan, and J.~Han.
\newblock Mining graph patterns efficiently via randomized summaries.
\newblock {\em Proc. VLDB Endow.}, 2(1):742--753, Aug. 2009.

\bibitem{Cheng2008}
J.~Cheng, J.~Yu, B.~Ding, P.~Yu, and H.~Wang.
\newblock Fast graph pattern matching.
\newblock In {\em Data Engineering, 2008. ICDE 2008. IEEE 24th International
  Conference on}, pages 913--922, April 2008.

\bibitem{Cheng2013}
J.~Cheng, X.~Zeng, and J.~Yu.
\newblock Top-k graph pattern matching over large graphs.
\newblock In {\em Data Engineering (ICDE), 2013 IEEE 29th International
  Conference on}, pages 1033--1044, April 2013.

\bibitem{William1995}
W.~J. Christmas, J.~Kittler, and M.~Petrou.
\newblock Structural matching in computer vision using probabilistic
  relaxation.
\newblock {\em IEEE Transactions on Pattern Analysis and Machine Intelligence -
  TPAMI}, 17(8):749--764, 1995.

\bibitem{Conte2004}
D.~Conte, P.~Foggia, C.~Sansone, and M.~Vento.
\newblock {Thirty Years of Graph Matching in Pattern Recognition}.
\newblock {\em International Journal of Pattern Recognition and Artificial
  Intelligence}, 18:265--298, 2004.

\bibitem{Cordella2004}
L.~P. Cordella, P.~Foggia, C.~Sansone, and M.~Vento.
\newblock {A (Sub)Graph Isomorphism Algorithm for Matching Large Graphs}.
\newblock {\em IEEE Transactions on Pattern Analysis and Machine Intelligence},
  26:1367--1372, 2004.

\bibitem{Cormen2001}
T.~H. Cormen, C.~E. Leiserson, R.~L. Rivest, and C.~Stein.
\newblock {\em {Introduction to Algorithms, Second Edition}}.
\newblock The MIT Press, 2001.

\bibitem{Eshera1984a}
M.~Eshera and K.~Fu.
\newblock A similarity measure between attributed relational graphs for image
  analysis.
\newblock In {\em 7th International Conference on Pattern Recognition}, pages
  75--77, 1984.

\bibitem{Eshera1984b}
M.~Eshera and K.-S. Fu.
\newblock A graph distance measure for image analysis.
\newblock {\em IEEE Transactions on Systems, Man and Cybernetics},
  SMC-14(3):398--408, May 1984.

\bibitem{Fan2010}
W.~Fan, J.~Li, S.~Ma, N.~Tang, Y.~Wu, and Y.~Wu.
\newblock Graph pattern matching: From intractable to polynomial time.
\newblock {\em Proc. VLDB Endow.}, 3(1-2):264--275, Sept. 2010.

\bibitem{Fan2012b}
W.~Fan, J.~Li, X.~Wang, and Y.~Wu.
\newblock Query preserving graph compression.
\newblock In {\em Proceedings of the 2012 ACM SIGMOD International Conference
  on Management of Data}, SIGMOD '12, pages 157--168, New York, NY, USA, 2012.
  ACM.

\bibitem{Fard2013}
A.~Fard, M.~U. Nisar, L.~Ramaswamy, J.~A. Miller, and M.~Saltz.
\newblock A distributed vertex-centric approach for pattern matching in massive
  graphs.
\newblock In {\em Proceedings of the 2013 IEEE International Conference on Big
  Data, 6-9 October 2013, Santa Clara, CA, USA}, pages 403--411, 2013.

\bibitem{Fard2014}
A.~Fard, M.~U. Nisar, L.~Ramaswamy, J.~A. Miller, and M.~Saltz.
\newblock Distributed and scalable graph pattern matching: Models and
  algorithms.
\newblock {\em Intenational Journal of Big Data}, 1(1):1--14, 2014.

\bibitem{Brian1939}
B.~Gallagher.
\newblock {\em Matching Structure and Semantics: A Survey on Graph-Based
  Pattern Matching}.
\newblock PhD thesis, 1939.

\bibitem{Gallagher2006}
B.~Gallagher.
\newblock Matching structure and semantics: A survey on graph-based pattern
  matching.
\newblock {\em AAAI FS}, 6:45--53, 2006.

\bibitem{Gallai1967}
T.~Gallai.
\newblock Transitiv orientierbare graphen.
\newblock {\em Acta Mathematica Hungarica}, 18:25--66, 1967.

\bibitem{Gao2010}
X.~Gao, B.~Xiao, D.~Tao, and X.~Li.
\newblock A survey of graph edit distance.
\newblock {\em Pattern Analysis Applications}, (13):113--129, 2010.

\bibitem{Garey1979}
M.~R. Garey and D.~S. Johnson.
\newblock {\em {Computers and Intractability: A Guide to the Theory of
  NP-Completeness}}.
\newblock 1979.

\bibitem{Gartner2003}
T.~Gartner, P.~Flach, and S.~Wrobel.
\newblock On graph kernels: Hardness results and efficient alternatives.
\newblock In Springer, editor, {\em Annual Conf. Computational Learning
  Theory}, pages 129--143, 2003.

\bibitem{Habib2010}
M.~Habib and C.~Paul.
\newblock A survey of the algorithmic aspects of modular decomposition.
\newblock {\em Computer Science Review}, 4(1):41--59, 2010.

\bibitem{Han2013}
W.-S. Han, J.~Lee, and J.-H. Lee.
\newblock Turboiso: Towards ultrafast and robust subgraph isomorphism search in
  large graph databases.
\newblock In {\em Proceedings of the 2013 ACM SIGMOD International Conference
  on Management of Data}, SIGMOD '13, pages 337--348, New York, NY, USA, 2013.
  ACM.

\bibitem{Hart1968}
P.~Hart, N.~Nilsson, and B.~Raphael.
\newblock {A Formal Basis for the Heuristic Determination of Minimum Cost
  Paths}.
\newblock {\em IEEE Transactions on Systems Science and Cybernetics},
  4:100--107, 1968.

\bibitem{Haussler1999}
D.~Haussler.
\newblock Convolution kernels on discrete structures.
\newblock Technical Report UCSC-CRL-99-10, University of California, Santa
  Cruz, 1999.

\bibitem{Henzinger1995}
M.~R. Henzinger, T.~A. Henzinger, and P.~W. Kopke.
\newblock {Computing Simulations on Finite and Infinite Graphs}.
\newblock In {\em IEEE Symposium on Foundations of Computer Science}, pages
  453--462, 1995.

\bibitem{Jouili2009}
S.~Jouili and S.~Tabbone.
\newblock Attributed graph matching using local descriptions.
\newblock In {\em ACIVS 2009, LNCS 5807}, pages 89--99, 2009.

\bibitem{Khan2011}
A.~Khan, N.~Li, X.~Yan, Z.~Guan, S.~Chakraborty, and S.~Tao.
\newblock Neighborhood based fast graph search in large networks.
\newblock In {\em Proceedings of the ACM SIGMOD International Conference on
  Management of Data, SIGMOD 2011, Athens, Greece, June 12-16}, pages 901--912,
  2011.

\bibitem{Khan2013}
A.~Khan, Y.~Wu, C.~C. Aggarwal, and X.~Yan.
\newblock Nema: Fast graph search with label similarity.
\newblock {\em PVLDB}, 6(3):181--192, 2013.

\bibitem{Khoo2001}
K.~G. Khoo and P.~N. Suganthan.
\newblock Multiple relational graphs mapping using genetic algorithms.
\newblock pages 727--737, 2001.

\bibitem{Kuhn1955}
H.~W. Kuhn.
\newblock {The Hungarian method for the assignment problem}.
\newblock {\em Naval Research Logistics Quarterly}, 2:83--97, 1955.

\bibitem{Lagraa2014}
S.~Lagraa, H.~Seba, A.~M'Baya, R.~Khennoufa, and H.~Kheddouci.
\newblock {A Distance Measure for Large Graphs based on Prime Graphs}.
\newblock {\em Pattern Recognition}, 2013.

\bibitem{Lopresti2001}
D.~P. Lopresti and G.~T. Wilfong.
\newblock {Comparing Semi-Structured Documents via Graph Probing}.
\newblock In {\em Workshop on Multimedia Information Systems}, pages 41--50,
  2001.

\bibitem{Ma2011}
S.~Ma, Y.~Cao, W.~Fan, J.~Huai, and T.~Wo.
\newblock Capturing topology in graph pattern matching.
\newblock {\em Proc. VLDB Endow.}, 5(4):310--321, Dec. 2011.

\bibitem{Ma2012}
S.~Ma, Y.~Cao, J.~Huai, and T.~Wo.
\newblock Distributed graph pattern matching.
\newblock In {\em Proceedings of the 21st International Conference on World
  Wide Web}, WWW '12, pages 949--958, New York, NY, USA, 2012. ACM.

\bibitem{Malewicz2009}
G.~Malewicz, M.~H. Austern, A.~J. Bik, J.~C. Dehnert, I.~Horn, N.~Leiser, and
  G.~Czajkowski.
\newblock Pregel: A system for large-scale graph processing - "abstract".
\newblock In {\em Proceedings of the 28th ACM Symposium on Principles of
  Distributed Computing}, PODC '09, pages 6--6, New York, NY, USA, 2009. ACM.

\bibitem{Mckay1981}
B.~McKay.
\newblock Practical graph isomorphism.
\newblock {\em Congress Numerantium}, 87:30--45, 1981.

\bibitem{Messmer1995}
B.~Messmer.
\newblock {\em Efficient Graph Matching Algorithms for Preprocessed Model
  Graphs}.
\newblock PhD thesis, University of Bern, Switzerland, 1995.

\bibitem{Messmer1999}
B.~T. Messmer and H.~Bunke.
\newblock A decision tree approach to graph and subgraph isomorphism detection.
\newblock {\em IEEE Transactions on Pattern Analysis and Machine Intelligence -
  TPAMI}, 32(12):1979--1998, 1999.

\bibitem{Micheli2009}
A.~Micheli.
\newblock Neural network for graphs : A contextual constructive approach.
\newblock {\em IEEE Transactions on Neural Networks}, 20(3):498--511, 2009.

\bibitem{Milner1989}
R.~Milner.
\newblock {\em Communication and Concurrency}.
\newblock Prentice Hall, 1989.

\bibitem{Munkres1957}
J.~Munkres.
\newblock Algorithms for the assignment and transportation problems.
\newblock {\em Journal of the Society for Industrial and Applied Mathematics},
  5:32--38, 1957.

\bibitem{Myers2000}
R.~Myers, R.~C. Wilson, and E.~R. Hancock.
\newblock Bayesian graph edit distance.
\newblock {\em IEEE Transactions on Pattern Analysis and Machine Intelligence -
  TPAMI}, 22(6):628--635, 2000.

\bibitem{Neuhaus2004}
M.~Neuhaus and H.~Bunke.
\newblock {An Error-Tolerant Approximate Matching Algorithm for Attributed
  Planar Graphs and Its Application to Fingerprint Classification}.
\newblock In {\em International Workshop on Structural and Syntactic Pattern
  Recognition}, pages 180--189, 2004.

\bibitem{Neuhaus2006c}
M.~Neuhaus and H.~Bunke.
\newblock {A Random Walk Kernel Derived from Graph Edit Distance}.
\newblock In {\em International Workshop on Structural and Syntactic Pattern
  Recognition}, pages 191--199, 2006.

\bibitem{Neuhaus2006a}
M.~Neuhaus and H.~Bunke.
\newblock A convolution edit kernel for error-tolerant graph matching.
\newblock In {\em IEEE international conference on pattern recognition, Hong
  Kong}, pages 220--223, 2006.

\bibitem{Neuhaus2007}
M.~Neuhaus and H.~Bunke.
\newblock {Automatic learning of cost functions for graph edit distance}.
\newblock {\em Information Sciences}, 177:239--247, 2007.

\bibitem{Pearl1982}
J.~Pearl.
\newblock Reverend bayes on inference engines: a distributed hierarchical
  approach.
\newblock In {\em in Proceedings of the National Conference on Artificial
  Intelligence}, pages 133--136, 1982.

\bibitem{Raveaux2010}
R.~Raveaux, J.-C. Burie, and J.-M. Ogier.
\newblock {A graph matching method and a graph matching distance based on
  subgraph assignments}.
\newblock {\em Pattern Recognition Letters}, 31:394--406, 2010.

\bibitem{Raymond2002}
J.~W. Raymond, E.~J. Gardiner, and P.~Willett.
\newblock {RASCAL: Calculation of Graph Similarity using Maximum Common Edge
  Subgraphs}.
\newblock {\em The Computer Journal}, 45:631--644, 2002.

\bibitem{Riesen2009}
K.~Riesen and H.~Bunke.
\newblock Approximate graph edit distance computation by means of bipartite
  graph matching.
\newblock {\em Image and Vision Computing}, 27:950--959, 2009.

\bibitem{Riesen2007}
K.~Riesen, M.~Neuhaus, and H.~Bunke.
\newblock Bipartite graph matching for computing the edit distance of graph.
\newblock pages 1--12, 2007.

\bibitem{Robles2005}
A.~Robles-kelly and E.~R. Hancock.
\newblock {Graph Edit Distance from Spectral Seriation}.
\newblock {\em IEEE Transactions on Pattern Analysis and Machine Intelligence},
  27:365--378, 2005.

\bibitem{hancock2007}
A.~Robles-kelly and E.~R. Hancock.
\newblock {A Riemannian approach to graph embedding}.
\newblock {\em Pattern Recognition -PR}, 40(3):1042--1056, 2007.

\bibitem{Alberto2000}
A.~Sanfeliu, R.~Alqu\'{e}zar, and F.~Serratosa.
\newblock Clustering of attributed graphs and unsupervised synthesis of
  function-described graphs.
\newblock volume~2, pages 6022--6025, 2000.

\bibitem{Sanfeliu1983}
A.~Sanfeliu and K.~Fu.
\newblock A distance measure between attributed relational graphs for pattern
  recognition.
\newblock {\em IEEE Transactions on Systems, Man, and Cybernetics (Part B)},
  13(3):353--363, 1983.

\bibitem{Shapiro1985}
L.~G. Shapiro and R.~M. Haralick.
\newblock A metric for comparing relational descriptions.
\newblock {\em IEEE Trans. Pattern Anal. Mach. Intell.}, 7(1):90--94, 1985.

\bibitem{Suganthan2002}
P.~N. Suganthan.
\newblock Structural pattern recognition using genetic algorithms.
\newblock {\em Pattern Recognition - PR}, 35(9):1883--1893, 2002.

\bibitem{Sun2012}
Z.~Sun, H.~Wang, H.~Wang, B.~Shao, and J.~Li.
\newblock Efficient subgraph matching on billion node graphs.
\newblock {\em PVLDB}, 5(9):788--799, 2012.

\bibitem{Sutinen1995}
E.~Sutinen and J.~Tarhio.
\newblock On using q-gram locations in approximate string matching.
\newblock In P.~Spirakis, editor, {\em Algorithms-ESA'95}, volume 979 of {\em
  Lecture Notes in Computer Science}, pages 327--340. Springer Berlin
  Heidelberg, 1995.

\bibitem{Tong2007}
H.~Tong, C.~Faloutsos, B.~Gallagher, and T.~Eliassi-Rad.
\newblock Fast best-effort pattern matching in large attributed graphs.
\newblock In {\em Proceedings of the 13th ACM SIGKDD International Conference
  on Knowledge Discovery and Data Mining}, KDD '07, pages 737--746, New York,
  NY, USA, 2007. ACM.

\bibitem{Ullmann76}
J.~R. Ullmann.
\newblock {An Algorithm for Subgraph Isomorphism}.
\newblock {\em J. ACM}, 23(1):31--42, Jan. 1976.

\bibitem{Umeyama1988}
S.~Umeyama.
\newblock An eigen decomposition approach to wighted graph mathcing problems.
\newblock {\em IEEE Transactions on Pattern Analysis and Machine Intelligence -
  TPAMI}, 10(5):695--703, 1988.

\bibitem{Valiant1990}
L.~G. Valiant.
\newblock A bridging model for parallel computation.
\newblock {\em Commun. ACM}, 33(8):103--111, Aug. 1990.

\bibitem{Vento2013}
M.~Vento.
\newblock {A One Hour Trip in the World of Graphs, Looking at the Papers of the
  Last Ten Years}.
\newblock In W.~G. Kropatsch, N.~M. Artner, Y.~Haxhimusa, and X.~Jiang,
  editors, {\em Graph-Based Representations in Pattern Recognition}, volume
  7877 of {\em Lecture Notes in Computer Science}, pages 1--10. Springer Berlin
  Heidelberg, 2013.

\bibitem{Wallis2001}
W.~Wallis, P.~Shoubridge, M.~Kraetz, and D.~Ray.
\newblock Graph distances using graph union.
\newblock {\em Pattern Recognition Letters}, 22(6-7):701 -- 704, 2001.

\bibitem{Wang2012}
G.~Wang, B.~Wang, X.~Yang, and G.~Yu.
\newblock Efficiently indexing large sparse graphs for similarity search.
\newblock {\em Knowledge and Data Engineering, IEEE Transactions on},
  24(3):440--451, March 2012.

\bibitem{Wang2012b}
X.~Wang, X.~Ding, A.~K.~H. Tung, S.~Ying, and H.~Jin.
\newblock An efficient graph indexing method.
\newblock In {\em Proceedings of the 2012 IEEE 28th International Conference on
  Data Engineering}, ICDE '12, pages 210--221, Washington, DC, USA, 2012. IEEE
  Computer Society.

\bibitem{Xiao08}
Y.~Xiao, H.~Dong, W.~Wu, M.~Xiong, W.~Wang, and B.~Shi.
\newblock {Structure-based graph distance measures of high degree of
  precision}.
\newblock {\em Pattern Recognition}, 41:3547--3561, 2008.

\bibitem{Zeng2009}
Z.~Zeng, A.~K.~H. Tung, J.~Wang, J.~Feng, and L.~Zhou.
\newblock Comparing stars: On approximating graph edit distance.
\newblock {\em Proceedings of The Vldb Endowment - PVLDB}, 2(1):25--36, 2009.

\bibitem{Zhao2013}
X.~Zhao, C.~Xiao, X.~Lin, Q.~Liu, and W.~Zhang.
\newblock A partition-based approach to structure similarity search.
\newblock {\em Proc. VLDB Endow. PVLDB}, 7(3):169--180, 2013.

\bibitem{Zhao2012}
X.~Zhao, C.~Xiao, X.~Lin, and W.~Wang.
\newblock {Efficient Graph Similarity Joins with Edit Distance Constraints}.
\newblock In {\em IEEE 28th International Conference on Data Engineering (ICDE
  2012), Washington, DC, USA (Arlington, Virginia), 1-5 April}, pages 834--845,
  2012.

\bibitem{Zheng2013}
W.~Zheng, L.~Zou, X.~Lian, D.~Wang, and D.~Zhao.
\newblock Graph similarity search with edit distance constraint in large graph
  databases.
\newblock In {\em 22nd ACM International Conference on Information and
  Knowledge Management, CIKM'13, San Francisco, CA, USA, October 27-November 1,
  2013}, pages 1595--1600, 2013.

\bibitem{Zou2009}
L.~Zou, L.~Chen, and M.~T. \"{O}zsu.
\newblock Distance-join: Pattern match query in a large graph database.
\newblock {\em Proc. VLDB Endow.}, 2(1):886--897, Aug. 2009.

\end{thebibliography}

\end{document}